\RequirePackage{fix-cm}
\documentclass[aip,reprint]{revtex4-1}
\usepackage{graphicx}
\usepackage{amsmath,bm}
\usepackage{fixcmex}
\usepackage{hyperref}
\usepackage{xcolor}
\usepackage{siunitx}
\usepackage[capitalize]{cleveref}

\newcommand*\diff{\mathop{}\!\mathrm{d}}
\newcommand{\ld}{\lambda_{\mathrm{D}}}

\usepackage{tikz}
\usetikzlibrary{arrows.meta,decorations.markings}

\definecolor{ionred}{RGB}{255,10,10}
\definecolor{ionblue}{RGB}{20,176,225}
\definecolor{vectorblue}{RGB}{13,105,145}
\definecolor{labelorange}{RGB}{195,70,0}

\begin{document}

\title{Modeling electrolytes in nanopores by Monte Carlo simulations and the Bazant--Storey--Kornyshev model}

\author{Nader Nekoubin}
 \affiliation{Department of Physics, Norwegian University of Life Sciences, \AA s, Norway}
 \author{David Fertig}
 \email{david.fertig@nmbu.no}
 \affiliation{Department of Physics, Norwegian University of Life Sciences, \AA s, Norway}
\author{Dezs\H{o} Boda}
\affiliation{Center of Natural Science, University of Pannonia, Veszpr\'{e}m, Hungary}
 \author{Mathijs Janssen}
 \email{mathijs.a.janssen@nmbu.no}
 \affiliation{Department of Physics, Norwegian University of Life Sciences, \AA s, Norway}

\date{\today}

\begin{abstract}
We study electrolyte-filled cylindrical nanopores through Monte Carlo (MC) simulations and the Bazant--Storey--Kornyshev model, for different ionic valencies, sizes, and bulk concentrations, pore radii and surface charge densities.
Our model accounts for finite ion size through a Stern layer and we use de~Souza and Bazant's mechanical equilibrium principle to derive a boundary condition for the outer Helmholtz plane.
For 1:1 electrolytes and moderate surface charge densities, we find that both the classical Poisson--Boltzmann--Stern (PB--Stern) and our Bazant--Storey--Kornyshev--Boltzmann--Stern (BSKB--Stern) model fit MC data well---for 2:1 and 3:1 electrolytes, the BSKB--Stern outperforms the PB-Stern model.
Conversely, the BSKB--Stern model poorly fits MC data in other scenarios.
First, BSKB--Stern does not capture drying and apparent like-charge attraction in 3:1 electrolytes and small surface charge densities.
Second, BSKB--Stern only fits MC data for finely-tuned ionic diameters; for other ionic diameters, the ionic charge densities show oscillations or extended near-surface regions caused by ionic packing and strong Coulomb interactions, not captured by the BSKB--Stern model.
\end{abstract}

\maketitle

\section{Introduction}
Nanopores are cylindrical pores with at least one characteristic dimension between $\sim\SI{1}{\nano\meter}$ and $\SI{100}{\nano\meter}$~\cite{schoch2008transport,faucher2019critical}.
In nature, nanopores transport ions, water, and biomolecules, thus enabling key processes in organisms~\cite{doyle_science_1998,sather_arp_2003}, such as muscle movement~\cite{fill_pr_2002}.
Synthetic nanopores were developed to desalinate seawater~\cite{lynch2020water}, harvest energy~\cite{sirkin2020transport}, mimic diode-like ionic transport~\cite{ai2010effects}, generate blue energy~\cite{nekoubin2024highly}, purify gas and water~\cite{faucher2019critical}, sense (bio)molecules~\cite{howorka_csr_2009,vlassiouk_jacs_2009,mayer2022biological}, and for the detection and diagnosis applications in human health~\cite{bhatti2021recent}.
Underlying these applications is the pores' ability to rectify ionic currents~\cite{bocquet_csr_2009}, respond as a transistor~\cite{madai2018controlling}, sense molecules~\cite{sirkin2020transport}, or selectively let ions or solvent pass~\cite{lynch2020water}.
In turn, these responses rely on properties of the pore (geometry, distribution of surface charge and reactive sites), the electrolyte (charge number, particle sizes, diffusitivies, permittivity), and the driving mechanism for electrolyte flow (pressure, temperature, concentration, or electric potential gradients) \cite{dietzel2017flow,peters2016analysis,malgaretti2019driving}.

Electrolyte flow through nanopores has been studied through molecular dynamics \cite{thompson2003nonequilibrium} as well as continuum models such as the Poisson--Nernst--Planck (PNP) equation \cite{dietzel2017flow,peters2016analysis,malgaretti2019driving}.
For instance, molecular dynamics simulations~\cite{cruz2009ionic} were used to probe ionic conductance and current rectification.
While flow through nanopores is an inherently out-of-equilibrium process, in some cases, transport properties such as conductance can also be approximated through equilibrium simulations and models. 
This is because the electric field inside a nanopore (i.e., the driving force) can be considered constant for nanopores whose length is much larger than their radius. 
In turn, the flux density is proportional to the concentration, which can be approximated by equilibrium simulations~\cite{sarkadi2021nanotubes,sarkadi_jml_2022}.
Accordingly, the selectivity of nanopores was studied through Monte Carlo (MC) simulations and Poisson--Boltzmann (PB) theory~\cite{matejczyk_jcp_2017,valisko_jcp_2019}.
For small pore surface charge densities and monovalent electrolytes, PB theory predictions agree well with the MC simulations.
For large surface charges and multivalent electrolytes, the predictions of PB theory are worse.
The reasons behind the disagreement are the following:
First, PB theory treats ions as point particles, so it overpredicts ion concentrations near highly charged surfaces~\cite{kilic2007steric,borukhov1997steric,gillespie2015review}.
Second, PB theory misses strong electrostatic ion--ion correlations~\cite{levin2002electrostatic} that are dominating in multivalent electrolyte solutions, room-temperature ionic liquids, and molten salts~\cite{storey2012effects,santangelo2006computing,fertig_pccp_2020}.
Such correlations can cause overscreening, where the first layer of counterions next to a surface overcompensates the surface charge, followed by a layer with an excess of coions \cite{bazant_prl_2011,storey2012effects,valisko_aip_2018,gupta_prl_2020}.
Extensions to PB theory were developed to account for ion size~\cite{stern1924theory,borukhov1997steric,kilic2007steric,kornyshev2007double} and ion--ion correlations~\cite{bazant_prl_2011,santangelo2006computing,desouza_jpcc_2020,budkov2022electric}. 
While none of these extended models fully reproduce simulation data and generally perform worse than density functional theory (DFT) models~\cite{gillespie_jpcm_2005,valisko_aip_2018,gillespie_jcp_2021,hartel2017structure}, their appeal lies in their relatively easy and transparent mathematical structure, enabling physical interpretation.
Moreover, their relatively low computational cost enables rapid exploration of parameter spaces---within proven regimes of validity.

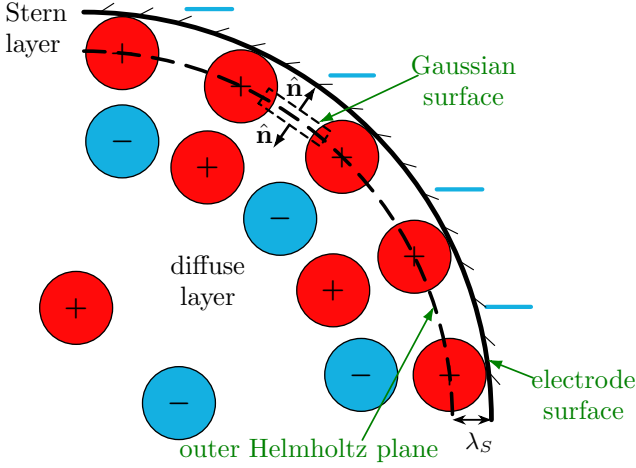
\begin{figure}
    \raggedright
    \begin{tikzpicture}[
    x=0.67cm,
    y=0.67cm,
    line cap=round,
    line join=round,
    every node/.style={
        font=\rmfamily\normalsize
    }
]

% ============================================================
% Bounding box
% ============================================================

\path[use as bounding box] (0,0) rectangle (12.90,10.20);

% ============================================================
% Common geometrical parameters
% ============================================================

% Centre of the circular curves
\def\xc{1.75}
\def\yc{1.75}

% Electrode-surface radius
\def\Relectrode{8.05}

% Common radius of every red and blue ion
\def\Rion{0.72}

% Radius through the centres of the first-layer ions
\def\Rhelmholtz{7.28}

% Radius through the centres of the second-layer ions
\def\Rsecond{5.55}

% Uniform first-layer angles
\def\aA{84}
\def\aB{64.75}
\def\aC{45.50}
\def\aD{26.25}
\def\aE{7}

% Dashed curve reaches exactly the horizontal radial line
\def\adashstart{90}
\def\adashend{0}

% Double-headed arrow located exactly at zero degrees
\def\agaparrow{0}

% ============================================================
% Second ion layer
% ============================================================

% Blue ions
\foreach \ang in {82.12,45.66,9.20}
{
    \pgfmathsetmacro{\xion}{\xc + \Rsecond*cos(\ang)}
    \pgfmathsetmacro{\yion}{\yc + \Rsecond*sin(\ang)}

    \filldraw[
        fill=ionblue,
        draw=black,
        line width=0.50pt
    ]
        (\xion,\yion) circle[radius=\Rion];

    \node[
        text=black
    ]
        at (\xion,\yion)
        {$\bm{-}$};
}

% Red ions
\foreach \ang in {63.89,27.43}
{
    \pgfmathsetmacro{\xion}{\xc + \Rsecond*cos(\ang)}
    \pgfmathsetmacro{\yion}{\yc + \Rsecond*sin(\ang)}

    \filldraw[
        fill=ionred,
        draw=black,
        line width=0.50pt
    ]
        (\xion,\yion) circle[radius=\Rion];

    \node[
        text=black
    ]
        at (\xion,\yion)
        {$\bm{+}$};
}

% ============================================================
% Bulk ions
% ============================================================

% One red bulk ion
\filldraw[
    fill=ionred,
    draw=black,
    line width=0.50pt
]
    (1.60,3.96) circle[radius=\Rion];

\node[
    text=black
]
    at (1.60,3.96)
    {$\bm{+}$};

% One blue bulk ion
\filldraw[
    fill=ionblue,
    draw=black,
    line width=0.50pt
]
    (3.63,2.08) circle[radius=\Rion];

\node[
    text=black
]
    at (3.63,2.08)
    {$\bm{-}$};

% ============================================================
% First ion layer adjacent to the electrode
% ============================================================

\foreach \ang in {\aA,\aB,\aC,\aD,\aE}
{
    \pgfmathsetmacro{\xion}{\xc + \Rhelmholtz*cos(\ang)}
    \pgfmathsetmacro{\yion}{\yc + \Rhelmholtz*sin(\ang)}

    \filldraw[
        fill=ionred,
        draw=black,
        line width=0.50pt
    ]
        (\xion,\yion) circle[radius=\Rion];

    \node[
        text=black
    ]
        at (\xion,\yion)
        {$\bm{+}$};
}

% ============================================================
% Fixed surface charges
% ============================================================

\draw[
    ionblue,
    line width=1.60pt
]
    (3.78,9.86) -- (4.68,9.86);

\draw[
    ionblue,
    line width=1.60pt
]
    (6.62,8.55) -- (7.48,8.55);

\draw[
    ionblue,
    line width=1.60pt
]
    (8.72,6.30) -- (9.62,6.30);

\draw[
    ionblue,
    line width=1.60pt
]
    (9.70,3.98) -- (10.58,3.98);

% ============================================================
% Electrode surface
% ============================================================

\draw[
    line width=1.80pt,
    postaction={decorate},
    decoration={
        markings,
        mark=between positions 0.025 and 0.975 step 0.060 with {
            \draw[
                line width=0.42pt
            ]
                (0,0) -- (0.27,0.19);
        }
    }
]
    (\xc,{\yc+\Relectrode})
    arc[
        start angle=90,
        end angle=0,
        radius=\Relectrode
    ];

% ============================================================
% Outer Helmholtz plane
% ============================================================

\draw[
    line width=1.35pt,
    dash pattern=on 8pt off 5pt
]
    ({\xc+\Rhelmholtz*cos(\adashstart)},
     {\yc+\Rhelmholtz*sin(\adashstart)})
    arc[
        start angle=\adashstart,
        end angle=\adashend,
        radius=\Rhelmholtz
    ];

% ============================================================
% Double-headed arrow between dashed and solid curves
% at exactly zero degrees
% ============================================================

\draw[
    <->,
    >={Stealth[length=1.8mm,width=1.3mm]},
    black,
    line width=0.70pt
]
    ({\xc+\Rhelmholtz*cos(\agaparrow)},
     {\yc+\Rhelmholtz*sin(\agaparrow)})
    --
    ({\xc+\Relectrode*cos(\agaparrow)},
     {\yc+\Relectrode*sin(\agaparrow)});

% Lambda_S label
\node[
    text=black,
    anchor=north,
    font=\rmfamily\fontsize{9pt}{12pt}\selectfont
]
    at ({\xc+0.5*(\Rhelmholtz+\Relectrode)+0.15},
        {\yc-0.08})
    {$\lambda_S$};

% ============================================================
% Rectangular element, n-vectors, and labels
% ============================================================

\begin{scope}[
    shift={({\xc+\Rhelmholtz*cos(55.125)},
            {\yc+\Rhelmholtz*sin(55.125)})},
    rotate=-34.875
]

    % Dashed black rectangle
    \draw[
        black,
        line width=0.85pt,
        dash pattern=on 3.1pt off 2.2pt
    ]
        (-0.74,-0.18)
        rectangle
        (0.74,0.18);

    % Attachment point on the upper long side
    \coordinate (gaussianLongSide) at (0.44,0.18);

    % Two short n-vectors
    \draw[
        -{Latex[length=2mm,width=1.3mm]},
        black,
        line width=1.00pt
    ]
        (0.00,-0.18) -- (0.00,-0.76);

    \draw[
        -{Latex[length=2mm,width=1.3mm]},
        black,
        line width=1.00pt
    ]
        (0.00,0.18) -- (0.00,0.76);

    % n labels
    \node
        at (-0.31,-0.60)
        {$\hat{\mathbf{n}}$};

    \node
        at (-0.35,0.50)
        {$\hat{\mathbf{n}}$};

\end{scope}

% ============================================================
% Text labels
% ============================================================

% Stern-layer label
\node[
    anchor=north west,
    align=left
]
    at (0.03,10.15)
    {Stern\\layer};

% Diffuse-layer label
\node[
    align=center
]
    at (4.20,4.45)
    {diffuse\\layer};

% ============================================================
% Outer-Helmholtz-plane label and arrow
% ============================================================

\draw[
    green!50!black,
    -{Latex[length=2mm,width=1.3mm]},
    line width=0.65pt
]
    (7.25,1.35)
    --
    ({\xc+\Rhelmholtz*cos(18)},
     {\yc+\Rhelmholtz*sin(18)});

\node[
    text=green!50!black
]
    at (6.20,1.15)
    {outer Helmholtz plane};

% ============================================================
% Electrode-surface label and arrow
% ============================================================

\draw[
    green!50!black,
    -{Latex[length=2mm,width=1.3mm]},
    line width=0.65pt
]
    (10.62,2.58)
    --
    ({\xc+\Relectrode*cos(8.5)},
     {\yc+\Relectrode*sin(8.5)});

\node[
    text=green!50!black,
    align=center
]
    at (11.62,2.25)
    {electrode\\surface};

% ============================================================
% Gaussian-surface label and arrow
% ============================================================

\draw[
    green!50!black,
    -{Latex[length=2mm,width=1.3mm]},
    line width=0.65pt
]
    (8.35,8.55)
    --
    (gaussianLongSide);

\node[
    text=green!50!black,
    align=center
]
    at (9.25,8.45)
    {Gaussian\\surface};

\end{tikzpicture}
    \centering
    \caption{Setup of the model, where finite ion size is captured through a Stern layer of width $\lambda_S$. Red and blue circles represent cations and anions, respectively, while the blue line segments represent fixed surface charges. The figure also indicates the Gaussian surface used in the derivation of \cref{eq:no_jump} and its unit normal outward vector, $\hat{\mathbf{n}}$.}
    \label{fig:BC}
\end{figure}

Here, we seek a more accurate continuum model for an electrolyte in a nanopore than the PB theory, which is still in use~\cite{ramirez_jms_2018,sarkadi_jml_2022}.
Specifically, we demarcate the validity of an extended Bazant--Storey--Kornyshev (BSK) model by comparing its predicted electrolyte structure to MC simulations.
While the PB equation can be derived from a free energy functional with an ideal-gas and a mean-field Coulombic term \cite{hartel2017structure}, BSK theory extends such a functional by another mean-field term of Landau--Ginzburg-type~\cite{bazant_prl_2011}, effectively putting a weaker energetic penalty on more curved potential profiles.
The resulting BSK equation equals the PB equation plus a fourth-order gradient term $\ell_c^2 \nabla^4$ [cf.~\cref{eq:bsk}], giving the BSK equation the flexibility to capture overscreening~\cite{bazant_prl_2011}.
The so-called correlation length $\ell_c $ is not predicted within the BSK model itself; to determine $\ell_c$'s dependence on system parameters, de Souza and Bazant fitted the BSK model (with $\ell_c$ the only fit parameter) to MC data of multivalent electrolytes near a flat electrode and outside a cylindrical electrode~\cite{desouza_jpcc_2020}.
Here, we consider a nanopore instead, that is, an electrolyte inside a cylinder. 
Our BSK model accounts for finite ion size simply through a Stern layer, and we apply the mechanical equilibrium principle of de Souza and Bazant~\cite{desouza_jpcc_2020} at the outer Helmholtz plane (OHP), the diffuse layer--Stern layer interface, see \cref{fig:BC}.
We fit the resulting Bazant--Storey--Kornyshev--Boltzmann--Stern model (BSKB--Stern, in the following) to MC data for a wide range of electrolyte concentrations, ionic valencies and sizes, surface charge densities, and pore radii, again using $\ell_c$ as the only free fit parameter.

This article is structured as follows.
\Cref{sec:model} introduces the setup, the MC simulation methods, the governing equations of the BSKB--Stern model, its boundary conditions, and the nondimensional form of the equations.
In \cref{results}, we present fits of the BSKB--Stern model to MC data for wide range of parameters and we discuss limitations of the model.
We present conclusions in \cref{conclusions}.

\section{Model}\label{sec:model}
\subsection{Setup}
\label{setup}

Consider an infinitely long cylindrical pore of radius $R$, filled with a binary electrolyte with an electric permittivity $\varepsilon$ and temperature $T$.
The pore is in diffuse contact with a reservoir at concentration $n_\infty$.
We denote the ionic valencies by $z_+$ for cations and $z_-$ for anions and, likewise, the stoichiometric coefficients by $\nu_+$ and $\nu_-$.
We describe the system using a cylindrical coordinate system.
As the nanopore is infinitely long and has rotational symmetry, the observables described below do not depend on the axial and azimuthal coordinates.

\subsection{Monte Carlo simulations}
\label{MC}
As we cannot simulate an infinitely long pore with MC, we consider a pore of length $L=\SI{100}{\nano\meter}$ with periodic boundary conditions applied in the axial direction, by which we effectively simulate an infinite system.
We emulate a surface charge density $\sigma$ by putting point charges on an equidistant square grid with a grid size between $0.01-\SI{0.025}{\nano\meter}$ on the cylinder's surface.
The ions are modeled using the Restricted Primitive Model (RPM) of electrolytes, with ionic diameters $d$, carrying point charges at their centers, in a continuum dielectric background of relative permittivity $\varepsilon_r=78.5$ and temperature $T=\SI{298.15}{\kelvin}$.

Each MC run consists of two separate simulations.
First, using the adaptive grand canonical MC method of Ref.~\cite{malasics2010efficient}, we find the excess chemical potential $\mu_{\mathrm{EX},i}$ of ionic species $i$, which yields an electrolyte with a desired bulk concentration $n_{\infty}$ and composition $\nu_i$ and $z_i$.
Second, using $\mu_{\mathrm{EX},i}$ as an input, a MC simulation of 50000 MC cycles is carried out in the grand-canonical ensemble, in which the cationic and anionic concentration profiles, $n_+(r)$ and $n_-(r)$, are sampled.
More details about this method can be found in Ref.~\cite{szarvas_aip_2024}.

\subsection{Bazant--Storey--Kornyshev--Boltzmann--Stern model}
\label{BSK}
\subsubsection{Governing equations}
\label{Equations}
The MC simulations described above naturally give rise to a charge-free layer of width $d/2$ next to the electrode surface---a Stern layer---as can been seen in all figures of \cref{results}.
To account for this Stern layer in our continuum models, we divide the pore in a diffuse part, $0<r<R-\lambda_S$, and a Stern layer, $R-\lambda_S<r<R$, with $\lambda_S=d/2$ being the width of the Stern layer and $r$ the radial coordinate.

The BSK theory revolves around a modified permittivity, $\hat{\varepsilon}=\varepsilon(1-\ell_c^2\nabla^2)$, with $\ell_c$ being a correlation length.
Using this $\hat{\varepsilon}$ in the electric displacement $\mathbf{D}=\hat{\varepsilon}\mathbf{E}$ and the relation $\mathbf{E}=-\nabla\psi$ between electric field $\mathbf{E}$ and potential $\psi$, and inserting all in Gauss's law in differential form, $\nabla \cdot \mathbf{D}=q$, we find the BSK equation~\cite{bazant_prl_2011}, 
\begin{align}\label{eq:bsk}
    -\varepsilon \nabla^2(1-\ell_c^2\nabla^2)\psi=q,
\end{align}
where $q=z_+en_++z_-en_-$ is the ionic charge density and $e$ is the elementary charge.

Considering Boltzmann weights for the ionic concentrations and writing~\cref{eq:bsk} in cylindrical coordinates yields, for the diffuse layer,
\begin{align}
    &-\partial^2_{r} \psi-\dfrac{1}{r}\partial_{r} \psi+\ell_c^2\left(\partial^4_{r}\psi+\dfrac{2}{r}\partial^3_{r} \psi-\dfrac{1}{r^2}\partial^2_{r} \psi+\dfrac{1}{r^3}\partial_{r} \psi\right) \notag\\
    &= \dfrac{en_\infty}{\varepsilon}\left[z_+ \nu_+\exp\left(-\dfrac{z_+e\psi}{kT}\right)+z_- \nu_-\exp\left(-\dfrac{z_-e\psi}{kT}\right)\right],\notag\\
    &\text{for}\qquad 0<r<R-\lambda_S,
    \label{eq:D_BSK}
\end{align}
where $k$ is the Boltzmann constant.

In the Stern layer there are no ions and thus $q=0$.
We assume that there are no ionic correlations either (\mbox{$\ell_c=\SI{0}{nm}$}), so Gauss's law reduces to the Laplace equation,
\begin{align}
    &\partial^2_{r} \psi+\dfrac{1}{r}\partial_{r} \psi=0, \qquad \text{for}\qquad R-\lambda_S <r<R.
    \label{eq:D_Laplace}
\end{align}
We refer to \cref{eq:D_BSK,eq:D_Laplace} as the BSKB--Stern model.

In absence of correlations ($\ell_c=\SI{0}{nm}$), \cref{eq:D_BSK} reduces to the PB equation,
\begin{align}
    &-\partial^2_{r} \psi-\dfrac{1}{r}\partial_{r} \psi \notag\\
    &= \dfrac{en_\infty}{\varepsilon}\left[z_+ \nu_+\exp\left(-\dfrac{z_+e\psi}{kT}\right)+z_- \nu_-\exp\left(-\dfrac{z_-e\psi}{kT}\right)\right],\notag\\
    &\text{for}\qquad 0<r<R-\lambda_S.
    \label{eq:D_PB}
\end{align}
We refer to \cref{eq:D_PB,eq:D_Laplace} as the PB--Stern model; we use this simpler model in some parts of \cref{results} as a point of reference for our discussion of the BSKB--Stern model.

\subsubsection{Boundary conditions}
\label{boundary conditions}
The fourth-order \cref{eq:D_BSK} and second-order \cref{eq:D_Laplace} require six boundary conditions. 
The first two boundary conditions follow from the symmetry at the center of the pore,
\begin{align}
    &\partial_{r} \psi|_{r=0}=0,\label{eq:symmetry_conditions1}\\
    &\partial^3_{r} \psi|_{r=0}=0.\label{eq:symmetry_conditions2}
\end{align}
The third boundary condition is the continuity of the electric potential at the OHP,
\begin{equation}\label{eq:continuity_of_electric_potential}
\psi|_{r=(R-\lambda_S)^-}=\psi|_{r=(R-\lambda_S)^+} ,   
\end{equation}
where negative and positive signs in $(R-\lambda_S)^{\pm}$ refer to, respectively, the left and right side of OHP. 

The fourth and fifth boundary conditions follow from Gauss's law~\cite{griffiths2023introduction}, $\oint_S \mathbf{D}\cdot \hat{\mathbf{n}}\, da=Q$, where $\hat{\mathbf{n}}$ is the outward normal vector to the Gaussian surface $\mathcal{S}$, enclosing the free charge $Q$.
For $\mathcal{S}$, we choose Gaussian pillboxes of side area $A$ centered either around the Stern--electrode interface or the OHP, see \cref{fig:BC}. 
For the Stern--electrode interface, $\int_{\mathrm{Stern}} \mathbf{D}\cdot\hat{\mathbf{n}}\,\diff a+\int_{\mathrm{electrode}} \mathbf{D}\cdot\hat{\mathbf{n}}\,\diff a=Q$, where $\hat{\mathbf{n}}=-\hat{\mathbf{r}}$ in the Stern layer and $\hat{\mathbf{n}}=\hat{\mathbf{r}}$ in the electrode.
Considering the electrode to be a perfect conductor ($\mathbf{D}|_{\mathrm{electrode}}=\mathbf{0}$), we find 
\begin{align}\label{eq:surf}
    -\mathbf{D}\cdot\hat{\mathbf{r}}|_{r=R}A=\sigma A.
\end{align}
As before, we assume that $\ell_c=\SI{0}{nm}$ in the charge-free Stern layer.
The displacement in the BSK theory, \mbox{$\mathbf{D}=\varepsilon(1-\ell_c^2\nabla^2)\mathbf{E}$}, thus simplifies to $\mathbf{D}=\varepsilon\mathbf{E}$ which, combined with~\cref{eq:surf}, gives the fourth boundary condition,
 \begin{align}
    \label{eq:constant_surface_charge}
    &\partial_r \psi|_{r=R}=\frac{\sigma}{\varepsilon}.
\end{align}
For a Gaussian pillbox around the uncharged OHP, $\int_{\mathrm{diffuse}} \mathbf{D}\cdot\hat{\mathbf{n}}\,\diff a+\int_{\mathrm{Stern}} \mathbf{D}\cdot\hat{\mathbf{n}}\,\diff a=0$, 
where $\hat{\mathbf{n}}=-\hat{\mathbf{r}}$ in the diffuse layer and $\hat{\mathbf{n}}=\hat{\mathbf{r}}$ in the Stern layer, giving
\begin{equation}
    -\hat{\mathbf{r}}\cdot\mathbf{D}|_{r=(R-\lambda_S)^-}+\hat{\mathbf{r}}\cdot\mathbf{D}|_{r=(R-\lambda_S)^+}=0.
\end{equation}
Using $\mathbf{D}|_{r=(R-\lambda_S)^-}=-\hat{\varepsilon}\nabla\psi|_{r=(R-\lambda_S)^-}$ and $\mathbf{D}|_{r=(R-\lambda_S)^+}=-\varepsilon\nabla\psi|_{r=(R-\lambda_S)^+}$ gives the fifth boundary condition,
 \begin{align}
    \label{eq:no_jump}
    &\partial_r \psi-\ell_c^2\left(\partial^3_{r}\psi+\dfrac{1}{r}\partial^2_{r}\psi - \dfrac{1}{r^2}\partial_{r}\psi\right)\Bigg|_{r=(R-\lambda_S)^-} \notag\\
    &=\partial_{r} \psi\Big|_{r=(R-\lambda_S)^+}.
\end{align}

The sixth boundary condition follows from mechanical equilibrium~\cite{bazant2009towards,desouza_jpcc_2020}. 
We consider a stationary fluid, so there is no viscous stress in our system. 
Thus, the total force density is $\mathbf{f}_{\mathrm{tot}}=-\nabla p+\mathbf{f}$, where $\nabla p$ is the hydrodynamic pressure gradient and $\mathbf{f}=-n_+\nabla\mu_+-n_-\nabla\mu_-$ is the thermodynamic force density.
Another assumption of the BSK theory is that the excess electrochemical potential, $\mu_{\mathrm{EX}}$, is zero.
Then, the mean-field electrochemical potential $\mu_i$ for a dilute electrolyte solution is given by
\begin{equation}
    \mu_i=\mu_i^0+kT \ln{\left(\frac{n_i}{n_\infty}\right)}+z_ie\psi,
    \label{eq:elecChemPoten}
\end{equation}
where $\mu_i^0$ is the reference chemical potential. 
We find $\mathbf{f}=-kT\nabla (n_++n_-)-q\nabla\psi$ using \cref{eq:elecChemPoten}, which can be rewritten as \cite{bazant2009towards}
\begin{equation}
    \mathbf{f}=-\nabla p_0-q\nabla\psi,
    \label{eq:force_equation}
\end{equation}
where $p_0$ is the osmotic pressure and $-q\nabla\psi$ is the electrostatic force density, yielding $\mathbf{f}_{\mathrm{tot}}=-\nabla(p+p_0)-q\nabla\psi$.
Following Ref. \cite{desouza_jpcc_2020}, we neglect the contribution of (osmotic) pressure in imposing mechanical equilibrium at the OHP, resulting in balance of only electrostatic forces ($\mathbf{F}_e$) exerted from two sides of the interface
\begin{equation}\label{eq:force_e}
    \mathbf{F}_e|_{r=(R-\lambda_S)^-}+\mathbf{F}_e|_{r=(R-\lambda_S)^+}=\mathbf{0}.
\end{equation}

In \cref{eq:force_equation}, $-q\nabla\psi$ is the volume force, but instead of calculating the volume forces, we calculate the electrostatic forces applied on OHP using the Maxwell stress tensor $\mathbf{\tau}_e$.
We rewrite \cref{eq:force_e} as $-\int_S(\hat{\mathbf{r}}\cdot\mathbf{\tau}_e|_{r=(R-\lambda_S)^-})\diff a=-\int_S(\hat{\mathbf{r}}\cdot\mathbf{\tau}_e|_{r=(R-\lambda_S)^+})\diff a$, giving 
\begin{equation}
    \hat{\mathbf{r}}\cdot\mathbf{\tau}_e|_{r=(R-\lambda_S)^-}=\hat{\mathbf{r}}\cdot\mathbf{\tau}_e|_{r=(R-\lambda_S)^+}\,.
    \label{eq:equatingForces}
\end{equation}

To calculate the electrostatic forces on both sides of interface and apply the mechanical equilibrium condition, we use the Maxwell stress tensor~\cite{desouza_jpcc_2020} 
\begin{widetext}
\begin{equation}
    \mathbf{\tau}_e=
    \left\{
    \begin{aligned}
        &\varepsilon\mathbf{E}\mathbf{E}-\dfrac{1}{2}\varepsilon\mathbf{E}^2\mathbf{I}+\varepsilon \ell_c^2\Big[(\mathbf{E}\cdot\nabla^2\mathbf{E})\mathbf{I}-\mathbf{E}(\nabla^2\mathbf{E})-(\nabla^2\mathbf{E})\mathbf{E}+\dfrac{1}{2}(\nabla\cdot\mathbf{E})^2\mathbf{I}\Big],\,\qquad &\text{for}&\quad r<R-\lambda_S,\\
        &\varepsilon\mathbf{E}\mathbf{E}-\dfrac{1}{2}\varepsilon\mathbf{E}^2\mathbf{I},\,\qquad &\text{for} &\quad r>R-\lambda_S,
    \end{aligned}
    \right.
    \label{eq:Maxwell stress tensor}
\end{equation}
\end{widetext}
which, inserted into \cref{eq:equatingForces}, gives
\begin{align}
    &\dfrac{1}{2}\left(\partial_r\psi\right)^2-\ell_c^2\left(\partial_r^3\psi+\frac{1}{r}\partial_r^2\psi-\frac{1}{r^2}\partial_r\psi\right)\partial_r\psi\notag\\
    &+\frac{1}{2}\ell_c^2\left(\partial_r^2\psi+\frac{1}{r}\partial_r\psi\right)^2\Bigg|_{r=(R-\lambda_S)^-}=\frac{1}{2}\left(\partial_r\psi\right)^2\Bigg|_{r=(R-\lambda_S)^+}.
    \label{eq:eq_Maxwell}
\end{align}
Inserting $\partial_r\psi|_{r=(R-\lambda_S)^+}$ from \cref{eq:no_jump} into \cref{eq:eq_Maxwell}, we find an equality of squares, giving 
 \begin{align}
    \label{eq:mechanical_equilibrium}
    &\ell_c\left(\partial_r^3\psi+\dfrac{1}{r}\partial^2_{r}\psi- \dfrac{1}{r^2}\partial_{r}\psi\right)\Bigg|_{r=(R-\lambda_S)^-} \notag\\
    &=\pm\left(\partial_r^2\psi+\frac{1}{r}\partial_r\psi\right)\Bigg|_{r=(R-\lambda_S)^-}.
\end{align}
Following Ref.~\cite{desouza_jpcc_2020}, we use the minus sign in the above equation, giving the sixth boundary condition.

In the PB--Stern model, we use \cref{eq:symmetry_conditions1,eq:continuity_of_electric_potential,eq:constant_surface_charge,eq:no_jump} as boundary conditions, with \cref{eq:no_jump} becoming
\begin{equation}
    \label{eq:no_jump_PB}
    \partial_r \psi\Big|_{r=(R-\lambda_S)^-} =\partial_{r} \psi\Big|_{r=(R-\lambda_S)^+}.  
\end{equation}
Note that~\cref{eq:symmetry_conditions2} is not needed to solve the PB--Stern model, and~\cref{eq:eq_Maxwell} does not give any new boundary condition after inserting \cref{eq:no_jump_PB} when $\ell_c=\SI{0}{\nano\meter}$.

\subsubsection{Nondimensional formulation and solution of \cref{eq:D_Laplace}}
\label{dimensionlessformulation}
We use the Debye length, $\lambda_D=\sqrt{\varepsilon kT/e^2S}$, with $S=\nu_+n_\infty z_+^2+\nu_-n_\infty z_-^2$, to scale all lengths: \mbox{$\tilde{r}=r/\lambda_D$}, \mbox{$\delta=\ell_c/\lambda_D$}, \mbox{$\xi=(R-\lambda_S)/\lambda_D$}, and \mbox{$\tilde{R}=R/\lambda_D$}.
Moreover, we introduce the dimensionless potential \mbox{$\tilde{\psi}=e\psi/kT$}, charge density $\tilde{q}=q/(Se)$, surface charge density $\tilde{\sigma}=\sigma/(\lambda_D S e)$, and valency parameter $\tilde{z}_\pm=\nu_\pm z_\pm/(\nu_+z_+^2+\nu_-z_-^2)$.
With these parameters and variables, we write the BSK \cref{eq:D_BSK} in nondimensional form as
\begin{align}
    &\delta^2\left(\partial^4_{\tilde{r}}\tilde{\psi}+\dfrac{2}{\tilde{r}}\partial^3_{\tilde{r}}\tilde{\psi}-\dfrac{1}{\tilde{r}^2}\partial^2_{\tilde{r}}\tilde{\psi}+\dfrac{1}{\tilde{r}^3}\partial_{\tilde{r}}\tilde{\psi}\right) - \partial^2_{\tilde{r}}\tilde{\psi}-\dfrac{1}{\tilde{r}}\partial_{\tilde{r}}\tilde{\psi}\notag\\
    &= \tilde{z}_+\exp(-z_+\tilde{\psi})+\tilde{z}_-\exp(-z_-\tilde{\psi}),\notag
    \label{eq:BSK}\\
    &\text{for}\qquad 0 <\tilde{r}<\xi,
\end{align}
the Laplace equation \eqref{eq:D_Laplace} as
\begin{align}
    &\partial^2_{\tilde{r}}\tilde{\psi}+\dfrac{1}{\tilde{r}}\partial_{\tilde{r}}\tilde{\psi}=0,\qquad \text{for}\qquad \xi <\tilde{r}<\tilde{R},
    \label{eq:Laplace}
\end{align}
and the boundary conditions \cref{eq:continuity_of_electric_potential,eq:symmetry_conditions1,eq:symmetry_conditions2,eq:constant_surface_charge,eq:no_jump,eq:mechanical_equilibrium} as
\begin{subequations}
\begin{align}
    \label{eq:BC_BSK3}
    &\partial_{\tilde{r}}\tilde{\psi}\Big|_{\tilde{r}=0}=0,\\  
    \label{eq:BC_BSK4}
    &\partial^3_{\tilde{r}}\tilde{\psi}\Big|_{\tilde{r}=0}=0,\\
    \label{eq:BC_Stern2}
    &\tilde{\psi}\Big|_{\tilde{r}=\xi^-}=\tilde{\psi}\Big|_{\tilde{r}=\xi^+},\\
    \label{eq:BC_Stern1}
    &\partial_{\tilde{r}}\tilde{\psi}\Big|_{\tilde{r}=\tilde{R}}=\tilde{\sigma},\\
    &\partial_{\tilde{r}} \tilde{\psi}-\delta^2\left(\partial^3_{\tilde{r}}\tilde{\psi}+\dfrac{1}{\tilde{r}}\partial^2_{\tilde{r}}\tilde{\psi} - \dfrac{1}{\tilde{r}^2}\partial_{\tilde{r}}\tilde{\psi}\right)\Big|_{\tilde{r}=\xi^-}
    =\partial_{\tilde{r}} \tilde{\psi}\Big|_{\tilde{r}=\xi^+},    \label{eq:BC_BSK1}\\
    &\delta\left(\partial^3_{\tilde{r}}\tilde{\psi}+\dfrac{1}{\tilde{r}}\partial^2_{\tilde{r}}\tilde{\psi} - \dfrac{1}{\tilde{r}^2}\partial_{\tilde{r}}\tilde{\psi}\right)\Big|_{\tilde{r}=\xi^-}
    = - \partial^2_{\tilde{r}}\tilde{\psi}-\dfrac{1}{\tilde{r}}\partial_{\tilde{r}}\tilde{\psi}\Big|_{\tilde{r}=\xi^-}.\label{eq:BC_BSK2}
\end{align}
\end{subequations}

Notably, \cref{eq:Laplace} with \cref{eq:BC_Stern2,eq:BC_Stern1} can be solved analytically for the potential in the Stern layer:
\begin{equation}
    \label{eq:analytical solution}
    \tilde{\psi}(\tilde{r})=\tilde{\sigma}\tilde{R}\ln{\frac{\tilde{r}}{\xi}}+\tilde{\psi}\Big|_{\tilde{r}=\xi}, \qquad \text{for}\qquad \xi<\tilde{r}<\tilde{R},
\end{equation}
where $\tilde{\psi}\big|_{\tilde{r}=\xi}$ is the electric potential at the OHP.

What remains is the fourth-order differential equation~\eqref{eq:BSK} subject to the unchanged boundary conditions \cref{eq:BC_BSK3,eq:BC_BSK4,eq:BC_BSK2}; inserting \cref{eq:analytical solution} into the last boundary condition \cref{eq:BC_BSK1} gives
\begin{equation}
    \partial_{\tilde{r}}\tilde{\psi}-\delta^2\left.\left(\partial_{\tilde{r}}^3\tilde{\psi}+\frac{1}{\tilde{r}}\partial_{\tilde{r}}^2\tilde{\psi}-\frac{1}{\tilde{r}^2}\partial_{\tilde{r}}\tilde{\psi}\right)\right\vert_{\tilde{r}=\xi^-}= \dfrac{\tilde{\sigma}\tilde{R}}{\xi}.\label{eq:BC_BSK1_2}
\end{equation}

\begin{figure}
    \includegraphics[width=0.95\linewidth]{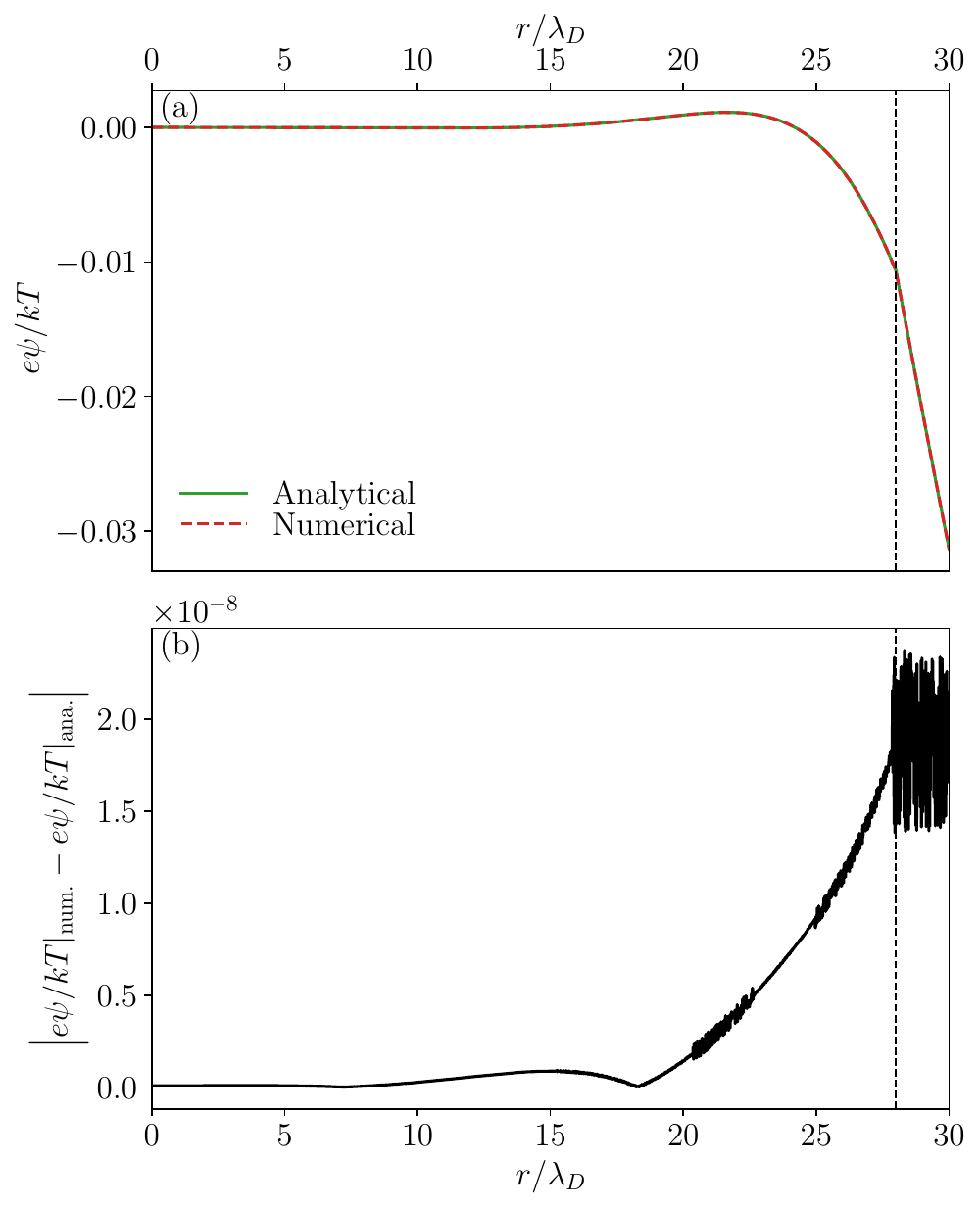}
    \caption{\label{fig:comparisonAnaNum} (a) Electric potential profiles for a 1:1 electrolyte, $\tilde{R}=30$, $ \xi=28, \tilde{\sigma}=-0.01, \mathrm{and}\, \delta=5$, calculated numerically (red dashed line) and analytically, by \cref{eq:analytical_solution} of the SI \cite{Note1} (green line). (b) The absolute difference between electric potentials calculated by the two methods.}
\end{figure}

For small applied surface charge, the potential is small throughout the pore ($\tilde{\psi}\ll1$) so that the Boltzmann weights in \cref{eq:BSK} can be linearized.
The resulting governing equations can then be solved analytically, as we show in \cref{Ana_solution} of the Supplemental Information (SI) \footnote{See the Supplemental Information for the analytical solution to the linearized BSKB--Stern model; details of its numerical implementation; MC data and BSKB--Stern fits to all considered parameter settings, and their corresponding $\delta$ fit values.\label{SIfootnote}}. 
For larger surface charges, the governing equations need to be solved numerically; we discuss the numerical implementation in \cref{numerical solution} of the SI \cite{Note1}.
After numerical solution of \cref{eq:BSK}, the electric potential in the diffuse layer and thus $\tilde{\psi}\big|_{\tilde{r}=\xi}$ are determined, so we can also complete our solution for the electric potential in the Stern layer [\cref{eq:analytical solution}].
\Cref{fig:comparisonAnaNum}(a) shows a characteristic potential profile $e\psi/kT$ vs. $r/\lambda_D$, obtained by the numerical and analytical methods, for a small surface charge density $\tilde{\sigma}=-0.01$, and parameters $\delta=5$, $\tilde{R}=30$, and $\xi=28$.
We observe excellent agreement between both methods---their difference is shown in \cref{fig:comparisonAnaNum}(b)---confirming the accuracy of the numerical solver.
The profile in \cref{fig:comparisonAnaNum}(a) is characteristic for an overscreening electrolyte: the small positive potential region around $r/\lambda_D=22$ is associated with a layer of charge of opposite sign as the charged ionic layer near the electrode.
Such overscreening profiles are visible for linearized BSK theory and a flat electrode for $\delta>1/2$~\cite{storey_pre_2012,fertig2025charging}.
For smaller $\delta$, the potential drops monotonically.

\section{Results}\label{results}

We performed MC simulations for 1:1, 2:1, 3:1, and 2:2 electrolytes with ion diameter $d=\SI{0.3}{\nano\meter}$, two bulk electrolyte concentrations ($n_\infty=0.1$ and $\SI{1}{M}$), two pore radii ($R=2$ and $\SI{5}{nm}$), and nine surface charge densities between $\sigma=0$ and $\SI{-3}{e/\nano\meter^2}$, and all combinations of these parameters.
Additionally, for the 2:2 and 3:1 electrolytes at $c=\SI{1}{M}$, $R=\SI{5}{\nano\meter}$, and $\sigma=\SI{-1}{e/\nano\meter^2}$, we varied the ion diameter between $d=0.15$ and $\SI{0.7}{\nano\meter}$.
In total, we simulated 159 simulation state points.
The above parameters fix the BSKB--Stern model, except the correlation parameter $\ell_c$.
For each state point, we fitted the BSKB--Stern model to the MC data using the Golden-Section search method, where $\delta=\ell_c/\lambda_D$ was varied over $\delta \in [0, 10]$ to minimize the root mean square difference between numerical and MC charge density profiles \cite{brent2013algorithms}.
In the SI \cite{Note1}, we show the MC data for 144 parameter settings with the same ion diameter, as well as the BSKB--Stern fits that converged within the given $\delta$ bracket (128 cases, not counting $\sigma=0$).
Here in the main text, we present several characteristic charge density profiles, and assess the validity and necessity of the BSKB--Stern model for them.

\begin{figure}
    \includegraphics[width=0.95\linewidth]{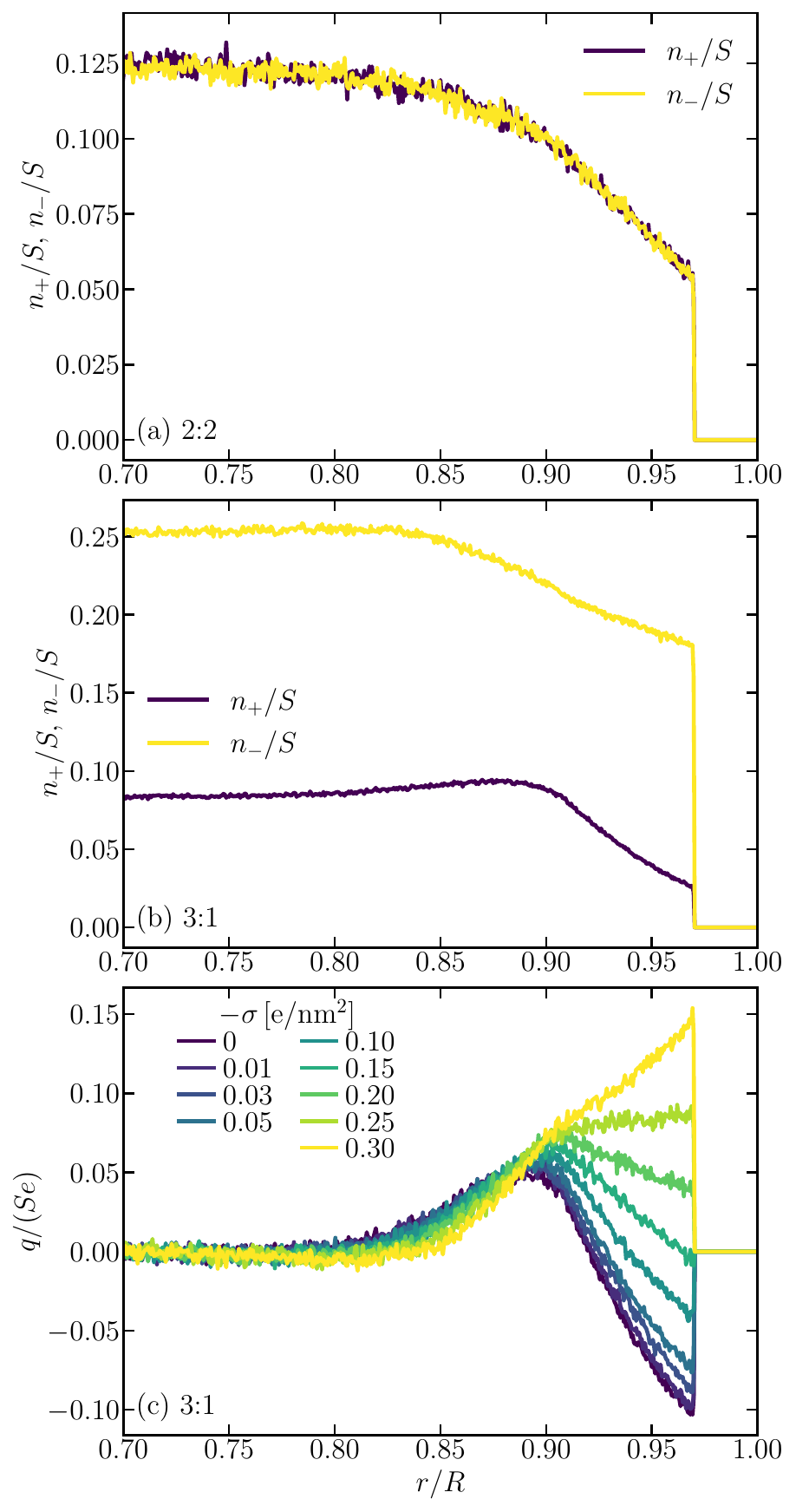}
    \caption{\label{fig:MC_chargeDensity_concen_S0} Cation (purple) and anion (yellow) concentration  profiles obtained from MC simulations for $R=\SI{5}{\nano\meter}$, $ n_\infty=\SI{1}{M}$, and $\sigma=\SI{0}{e/\nano\meter^2}$ for a 2:2 (a) and a 3:1 electrolyte (b). MC charge density profiles at $R=\SI{5}{\nano\meter}$ and $ n_\infty=\SI{1}{M}$ for various surface charge density values for a 3:1 electrolyte (c).}
\end{figure}

\begin{figure}
    \includegraphics[width=0.95\linewidth]{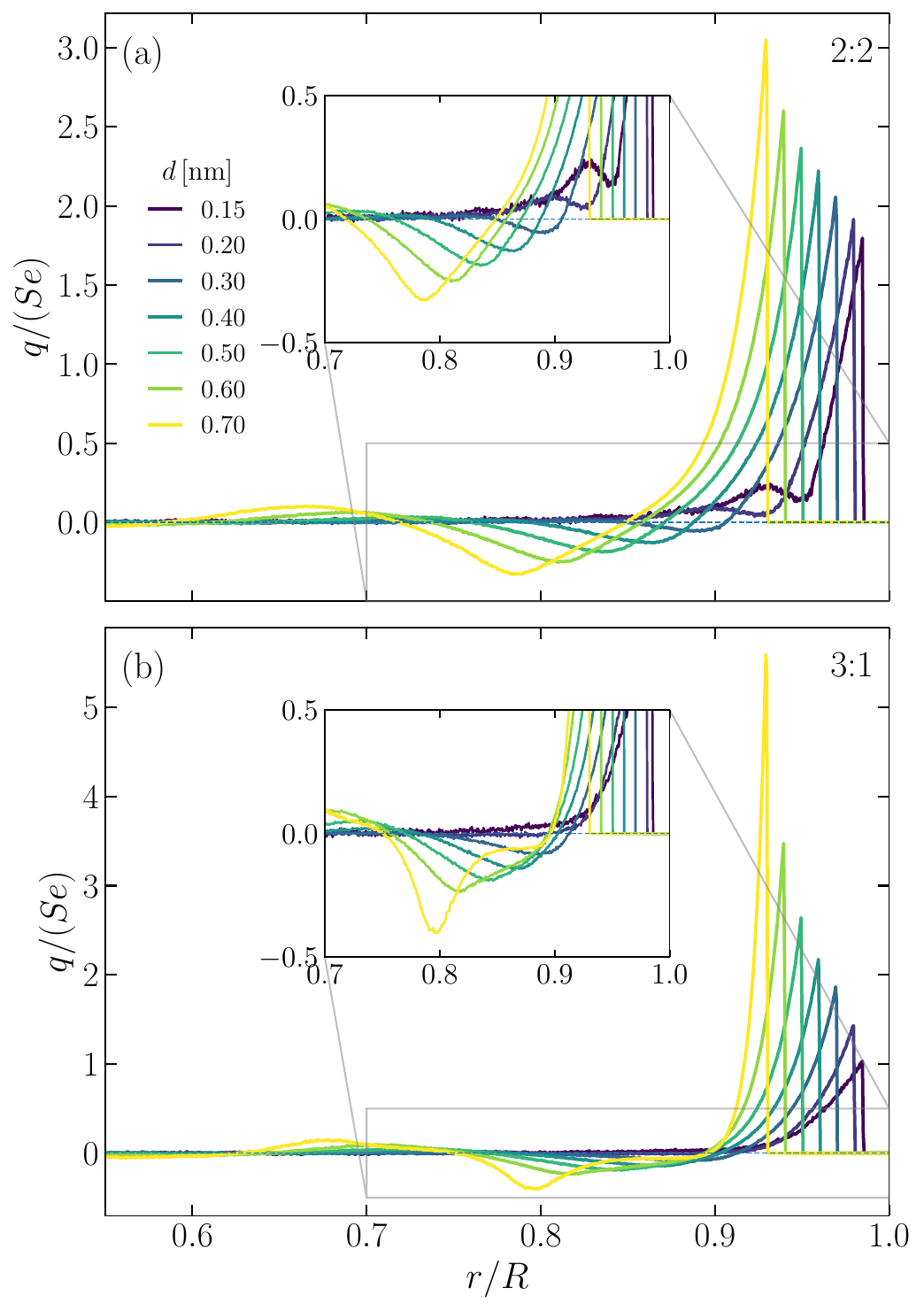}
    \caption{\label{fig:MC_chargeDensity_concen_S0_various_d} Ionic charge density profiles obtained from MC simulations at $R=\SI{5}{\nano\meter}$, $ n_\infty=\SI{1}{M}$, $\sigma=\SI{-1}{e/\nano\meter^2}$, and various ionic diameters between $d=0.15$ and $\SI{0.7}{\nano\meter}$ for a 2:2 (a) and a 3:1 electrolyte (b).}
\end{figure}

\subsection{Drying, like-charge-attraction, and packing not captured by the BSKB--Stern model}

\Cref{fig:MC_chargeDensity_concen_S0}(a) and (b) show MC results for the cationic and anionic densities profiles near an uncharged (\mbox{$\sigma=\SI{0}{e/\nano\meter^2}$}) pore wall for 2:2 (a) and 3:1 (b) electrolytes.
We see that neither electrolyte has a uniform concentration.
The electrolyte ``dries'' near the electrode: the concentration of ions is smaller near the wall than in the bulk.
For slit pores and symmetric electrolytes, such drying is predicted by the contact theorem of Henderson~\cite{henderson_jec_1979}.
For the symmetric 2:2 electrolyte, the cationic and anionic density profiles show the same monotonically decreasing trend; hence, their ionic charge density is uniformly zero.
For the valence-asymmetric 3:1 electrolyte, the ionic density profiles in \cref{fig:MC_chargeDensity_concen_S0}(b) again show wall-induced drying, but now also a cationic peak next to the depletion layer. 
Moreover, the densities now give rise to a nontrivial charge density, shown in \cref{fig:MC_chargeDensity_concen_S0}(c) with a purple line.
In line with previous literature~\cite{dicaprio_jpcb_2009}, we find that the ion with the higher valency is depleted more near the interface, leaving a surplus of the smaller valency ion---in our case, the anion---, which in turn induces a positive secondary layer [\cref{fig:MC_chargeDensity_concen_S0}(c), purple curve], where the cationic density has its weak maximum [\cref{fig:MC_chargeDensity_concen_S0}(b)]. 
Note that the pore is still overall electroneutral.
Beside the $\sigma=\SI{0}{e/\nano\meter^2}$ case discussed so far, \cref{fig:MC_chargeDensity_concen_S0}(c) shows MC results for the ionic charge densities for the same pore and 3:1 electrolyte, for eight small nonzero surface charge densities.
We observe negative contact values for the ionic charge density up to about $\sigma=\SI{-0.1}{e/\nano\meter^2}$, which seems to indicate like-charge attraction.
In reality, the anions are there despite the negative surface charge density, not because of it.
Drying is present at all surface charge values, although less at larger surface charge values.
As BSKB--Stern (or PB--Stern) cannot capture like charge attraction, it generally fitted the MC data for asymmetric electrolytes and small surface charge densities poorly, or the fitting procedure did not converge, see \cref{fig:S_charge_density_1,fig:S_charge_density_2,fig:S_charge_density_3,fig:S_charge_density_4} in the SI \cite{Note1}.

\Cref{fig:MC_chargeDensity_concen_S0_various_d} shows ionic density profiles for various ionic diameters between $d=0.15$ and $\SI{0.7}{\nano\meter}$.
For a 2:2 electrolyte [\cref{fig:MC_chargeDensity_concen_S0_various_d}(a)] and the two smallest ionic diameters, the ionic charge density shows a large peak, followed by a shoulder with the same sign.
With increasing ionic diameter, the shoulder becomes smaller, and, for $d=\SI{0.3}{\nano\meter}$, the ionic charge density shows a distinctive overscreening-type profile.
For larger $d$, we observe damped oscillations in the ionic charge density profile.
Notably, the BSKB--Stern model can only describe the intermediate behavior seen around $d=\SI{0.3}{\nano\meter}$---not the shoulder at small ionic diameters or the damped oscillations at larger $d$.
For the 3:1 electrolyte [\cref{fig:MC_chargeDensity_concen_S0_various_d}(b)], much of the same behavior is seen, though we do not observe the same shoulder for small ion diameters and the ionic charge density at contact increases stronger with increasing ion diameter.

Increasing ionic diameter introduces a competing effect, steric exclusion.
At a fixed concentration, the packing fraction $\eta=\frac{\pi}{6}\sum_i n_id_i^3$ increases with increasing ion diameter, and excluded-volume effects become more important.
Dense hard-sphere fluids also exhibit damped oscillatory number-density profiles near a hard interface~\cite{jagannathan_jcp_2002}.
In a system of charged hard spheres, this steric layering is coupled with electrostatics and therefore can produce oscillations in the ionic charge density profiles.
The oscillatory behavior increases with increasing ion diameter as observed in \cref{fig:MC_chargeDensity_concen_S0_various_d}.
Similar oscillatory decay can be observed in size asymmetric systems~\cite{valisko_aip_2018} when the trivalent cation is significantly larger than the monovalent ion.

Taken together, \cref{fig:MC_chargeDensity_concen_S0,fig:MC_chargeDensity_concen_S0_various_d} show that BSKB--Stern cannot be expected to describe the RPM in a nanopore if the surface charge density is low, as the BSK theory does not capture drying.
Moreover, EDL profiles are generally affected by the ion diameter (depending on surface charge densities and ion concentrations), and we only observed a BSK-like charge density profile for a particular parameter combination---for other parameters we saw damped oscillations or shoulders, which the BSK theory cannot capture.

\subsection{Nonoverscreening systems fitted well by BSKB--Stern}

\begin{figure}
    \includegraphics[width=0.95\linewidth]{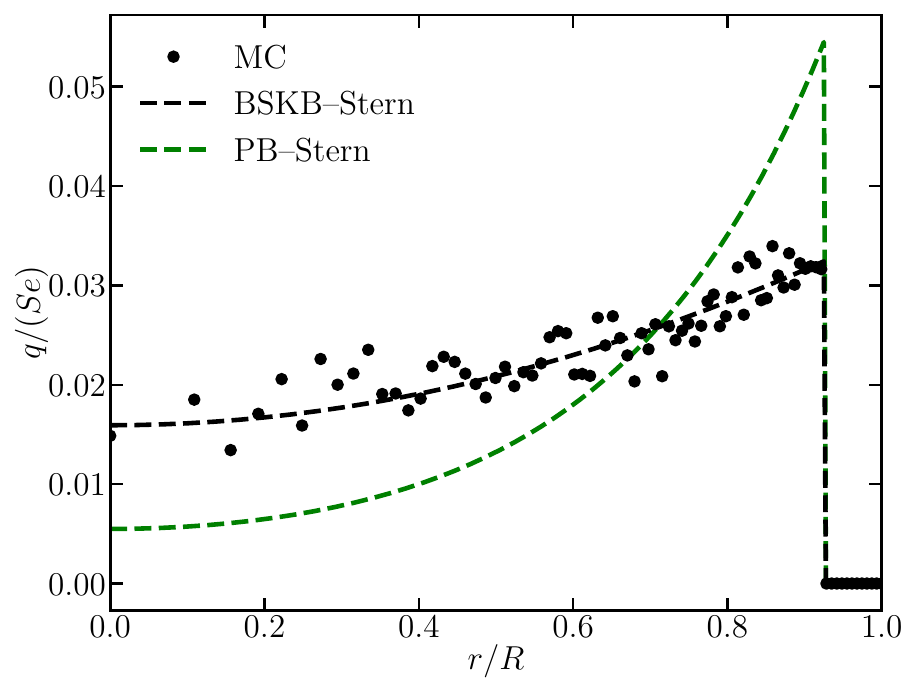}
    \caption{\label{fig:2_2 small charge} Charge density profiles for 2:2 electrolyte at $n_\infty=\SI{0.1}{M}$, $R=\SI{2}{\nano\meter}$, and $\sigma=\SI{-0.01}{e/\nano\meter^2}$, obtained from MC simulations (symbols) and numerical solutions based on the BSKB--Stern and PB--Stern models (dashed black and green lines).}
\end{figure}

\begin{figure*}
    \centering
    \includegraphics[width=0.95\textwidth]{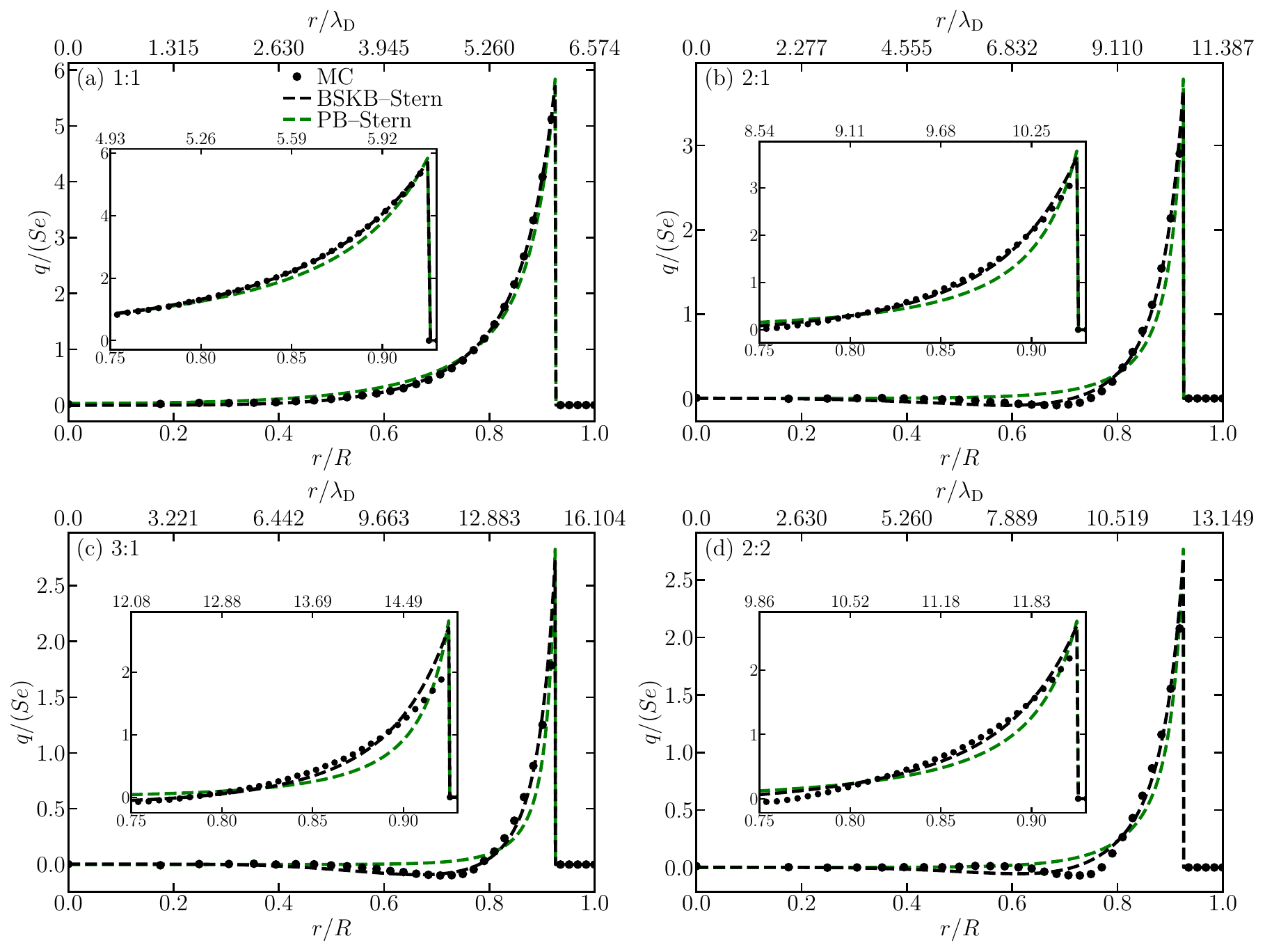}
    \caption{\label{fig:overscreening_capture_methods}
    Ionic charge density profiles for (a) 1:1, (b) 2:1, (c) 3:1, and (d) 2:2 electrolytes at
    $n_\infty=1\,\mathrm{M}$, $R=2\,\mathrm{nm}$, and $\sigma=-1\, \mathrm{e/nm^2}$ obtained by MC simulations (circles), BSKB--Stern (black dashed lines), and PB--Stern models (green dashed lines).}
\end{figure*}

\Cref{fig:2_2 small charge} shows the ionic charge density of a 2:2 electrolyte for $R=\SI{2}{\nano\meter}$ and $n_{\infty}=\SI{0.1}{M}$, and a small surface charge density of $\sigma=\SI{-0.01}{e/\nano\meter^2}$.
For these parameters, the MC data do not show negative charge density regions characteristic for overscreening.
This is not surprising, as overscreening requires strong ion-wall electrostatic correlations; hence, large $\sigma$.
Instead, we observe a nonzero charge density in the MC data in the middle of the pore ($r/R=0$), indicating double layer overlap; for the above parameters, $R/\ld=4.158$. 
We found that the BSKB--Stern model fits the MC data well for $\delta=9.67$.
Both BSKB--Stern model fit and the PB--Stern model reproduce the nonzero charge densities in the middle, but the PB--Stern model predicts stronger charge accumulation at the pore surface than observed in the MC data, with faster exponential decay.

For the mentioned $\delta$ fit value, the correlation length is almost ten times larger than the Debye length, significantly larger than the ion diameter, and also larger than the pore radius.
Still, in the MC simulations, ions can interact with each other (hence, be correlated) over long distances in axial direction.
Notably, $\delta$ values beyond $\sim R/\ld$ do not describe the physics the BSK model was devised for, that is: overscreening.
Still, the flexibility of the BSK equation, coming from its fourth order gradient and fitted correlation length, mean that the BSKB--Stern model can accurately fit nonoverscreening charge density profiles as well.

\subsection{Electrolytes and surface charge densities where BSKB--Stern fits MC data decently}
\Cref{fig:overscreening_capture_methods} shows ionic charge density profiles [$q/(S e)$ vs. $r/R$] obtained by MC simulations (black dots) and the BSKB--Stern (black dashed lines) and PB--Stern models (green dashed lines) for 1:1 (a), 2:1 (b), 3:1 (c), and 2:2 (d) electrolytes and a large surface charge density $\sigma=-1\, \mathrm{e/nm^2}$.
We only show a subset of the MC data for better visibility---the BSKB--Stern fits are performed based on the full dataset.
The insets of \cref{fig:overscreening_capture_methods} show the same data, zooming in near the electrode.
For the 1:1 electrolyte [\cref{fig:overscreening_capture_methods}(a)], the PB--Stern model predictions closely follow the MC data; the BSKB--Stern model follows them even closer.
In the MC data for the 2:1, 3:1, and 2:2 electrolytes shown in panels (b)-(d), the ionic charge densities are nonmonotonous, with a layer with a negative ionic charge density near the positive peaks---indicative for overscreening.
The PB--Stern model predicts monotonous density profiles; hence, cannot capture the negative $q$ region in the MC data.
The BSKB--Stern model performs somewhat better than PB--Stern, especially for the valence-asymmetric 2:1 and 3:1 electrolytes [\cref{fig:overscreening_capture_methods}(b) and (c)].
As mentioned, the BSK--Stern and PB-Stern theories treat Coulomb interactions in a mean-field manner; the error made in this approximation can be expected to increase with the ionic coupling parameter $\Gamma = z_+ |z_-| \lambda_B/d$~\cite{henderson_jctc_2009}, with $\lambda_B$ the Bjerrum length.
$\Gamma$ is the energy needed to bring two oppositely charged ions to contact scaled to the thermal energy.
For our systems, $\Gamma_{1:1}<\Gamma_{2:1}<\Gamma_{3:1}<\Gamma_{2:2}$.
Indeed we observe that the BSKB-Stern model fits the MC data from best to worst in that order.
In particular, the MC data show a sharper overscreening layer for the 2:2 electrolyte [\cref{fig:overscreening_capture_methods}(d)] than for the valence-asymmetric electrolytes, which the BSKB--Stern model cannot capture.

\begin{figure*}
    \centering
    \includegraphics[width=0.95\textwidth]{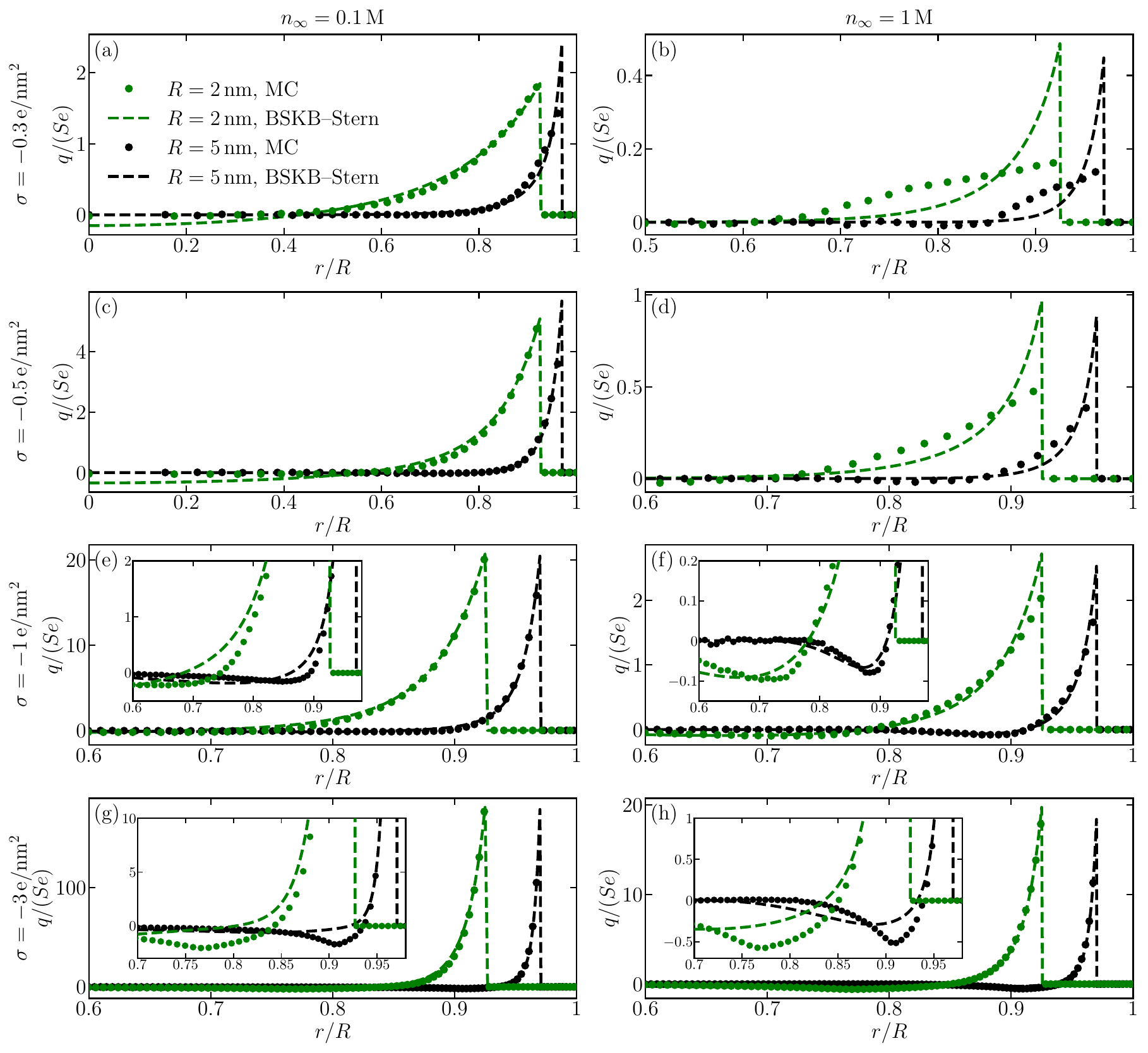}
    \caption{\label{fig:RadConcenCharge}
    Charge density profiles of 3:1 electrolyte at $n_\infty=0.1, \SI{1}{M}$ and $R=2, \SI{5}{\nano\meter}$ for various surface charges as indicated, obtained by the BSKB--Stern model and MC simulations.}
\end{figure*}

\Cref{fig:RadConcenCharge} reports charge density profiles for a 3:1 electrolyte calculated by MC simulations (dots) and the BSKB--Stern model (dashed lines) for two bulk concentrations ($n_\infty=\SI{0.1}{M}$, left column; and $\SI{1}{M}$, right column), two radii, ($R=2$ and $\SI{5}{\nano\meter}$), and large surface charge densities, $-\sigma=0.3,0.5,1$, and $\SI{3}{e/\nano\meter^2}$, corresponding to the four rows.
For $\sigma=\SI{-0.3}{e/\nano\meter^2}$, the BSKB--Stern model fits the MC simulations better for $n_{\infty}=\SI{0.1}{M}$ [\cref{fig:RadConcenCharge}(a)] than for $n_{\infty}=\SI{1}{M}$ [\cref{fig:RadConcenCharge}(b)].
In turn, \cref{fig:RadConcenCharge}(a) shows better fits for $R=\SI{5}{\nano\meter}$ than for $R=\SI{2}{\nano\meter}$, as the BSKB--Stern model overpredicts the EDL overlap in the middle of the nanopore.
In \cref{fig:RadConcenCharge}(b), for the concentrated solution ($n_{\infty}=\SI{1}{M}$), the MC charge density profiles still show the effect of wall-induced drying, not captured by the BSKB--Stern model [note that the black dots correspond to the same data as portrayed by the yellow line in \cref{fig:MC_chargeDensity_concen_S0} (c)].
For the slightly larger surface charge density $\sigma=\SI{-0.5}{e/\nano\meter^2}$ shown in \cref{fig:RadConcenCharge}(c) and (d), similar trends are visible:
better model fits to the MC data for the more dilute electrolyte ($n_{\infty}=\SI{0.1}{M}$) and a remnant of wall induced drying for the more concentrated electrolyte ($n_{\infty}=\SI{1}{M}$), though the model fits work better for $\sigma=\SI{-0.5}{e/\nano\meter^2}$ [\cref{fig:RadConcenCharge}(d)] than for $\sigma=\SI{-0.3}{e/\nano\meter^2}$ [\cref{fig:RadConcenCharge}(b)] as electrostatic interactions gradually overtake the drying effect [see also \cref{fig:RadConcenCharge}(f) and (h), where the drying is hardly visible].

Comparing all panels of \cref{fig:RadConcenCharge}, we find the best agreement between MC and the BSKB--Stern model for $\sigma=\SI{-1}{e/\nano\meter^2}$ [\cref{fig:RadConcenCharge}(e) and (f)], in particular for $n_{\infty}=\SI{1}{M}$.
Generally, the BSKB--Stern model predictions agree better with the MC data for larger systems (with smaller EDL overlap).
For $n_{\infty}=\SI{0.1}{M}$ (left column of~\cref{fig:RadConcenCharge}), the BSKB--Stern model captures the charge density peak near the OHP as well as the depth and width of the overscreening layer.
We observe that drying is less dominant for more dilute electrolytes (left column of~\cref{fig:RadConcenCharge}); it is balanced by smaller charge densities, therefore the BSKB--Stern model fits the MC data better.
In the center of the pore, especially for smaller pore radii [green curves in e.g. \cref{fig:RadConcenCharge}(a)], the two methods do not agree.
The cause of this might be that the electrostatic screening length is not the Debye length for the primitive model of electrolytes (see Ref.~\cite{gillespie_jcp_2021} for additional definitions of screening lengths of charged hard spheres), and the MC simulations predict a smaller overlap.

For smaller surface charges [$\sigma=\SI{-0.3}{e/\nano\meter^2}$ and $\SI{-0.5}{e/\nano\meter^2}$, \cref{fig:RadConcenCharge}(a)-(d)], the agreement is better for lower concentrations.
At $\sigma=\SI{-1,}{e/\nano\meter^2}$ [\cref{fig:RadConcenCharge}(e)-(f)], the two methods agree better for larger concentrations, where the BSKB--Stern model captures the width and depth of the overscreening layer.
For the large surface charge density $\sigma=\SI{-3}{e/\nano\meter^2}$ [\cref{fig:RadConcenCharge}(g)-(h)], for all shown state points, the two methods deviate more.
The charge density peak is captured, but the BSKB--Stern model predicts a wider overscreening layer than the MC simulations.

\section{Conclusions}
\label{conclusions}
We developed a BSKB--Stern model for ionic charge densities in nanopores and compared its predictions to MC simulations of the RPM for a wide range of ionic valencies, diameters, and bulk concentrations, and pore radii and surface charge densities.
Our model-versus-simulations study was similar to the work by de Souza and Bazant~\cite{desouza_jpcc_2020}, up to four differences.
First, we considered a different geometry---a cylindrical electrolyte-filled pore.
Second, instead of shifting the MC charge density profiles by one ion radius (see Supplementary Material of Ref.~\cite{desouza_jpcc_2020}), we explicitly included a Stern layer in the model.
Third, we applied the mechanical equilibrium at the OHP, while Ref.~\cite{desouza_jpcc_2020} applied it at the electrode--diffuse-layer interface.
Fourth, we considered a wider surface charge density range, four different electrolytes and ions with different diameters.

For 1:1 electrolytes, we found that the BSKB--Stern model fitted the MC data well, but so did the classical PB-Stern model.
For large surface charge densities, we observed progressively better fits of the BSKB--Stern model going from 2:2, 3:1, 2:1, to 1:1 electrolytes.
For these electrolytes, the ionic coupling parameter $\Gamma$ is progressively smaller.
Hence, the BSKB--Stern model fits MC data better the weaker the Coulomb interactions, as the error made in treating Coulomb interactions at mean-field level becomes progressively smaller.
For 3:1 electrolytes, we found that the BSKB--Stern model fitted the MC data well for many combinations of surface charge densities and bulk concentrations, except for dense electrolytes near weakly charged surfaces.
Moreover, for large surface charge densities, the BSKB--Stern model accurately described these 3:1 electrolytes' sharp first counterion layer, but not their subsequent coion layer.

While the BSK model was devised to describe overscreening, the flexibility offered by its fourth order gradient term and the fit parameter $\ell_c$ sometimes led to the BSKB--Stern model outperforming the PB--Stern model in cases with no overscreening.
For a 2:2 electrolyte and small surface charge density, for instance, the EDL in the nanopore were overlapping---we found good fits to the MC data for $\ell_c$ values roughly ten times the Debye length, and roughly two times the pore radius.

We also saw several parameter combinations for which the BSKB--Stern model was inadequate.
First, for asymmetric 2:1 and 3:1 electrolytes in pores with small surface charge densities, MC predicted charge density profiles that showed drying near the pore's surface, a beyond-mean-field effect not captured by the BSKB--Stern model.
Second, MC predicted charge density profiles typical for the BSKB--Stern model only for a particular ionic diameter $d=\SI{0.3}{\nano\meter}$.
For other ionic diameters, MC produced charge density profiles that the BSKB--Stern model cannot capture, with near-surface shoulders ($d<\SI{0.3}{\nano\meter}$) or oscillations ($d>\SI{0.3}{\nano\meter}$), caused by strong Coulombic and excluded volume interactions.
Notably, in their MC vs. BSK model comparison for the RPM near a flat wall, Ref.~\cite{desouza_jpcc_2020} considered only the $d=\SI{0.3}{\nano\meter}$ case---it would be worthwhile to see how the BSK model performs  for other ionic diameters.

In conclusion, the restricted range of parameters over which the BSKB--Stern model reproduces MC simulations of the RPM reflects its simplified mean-field description of finite-ion effects and Coulombic interactions.
While we considered the RPM in a nanopore here, the same conclusion can be expected to hold for the Primitive Model ($d_+\neq d_-$) and other geometries, in particular, near flat plates.

 \section*{Acknowledgements}
We thank Adrian L. Usler for inspiring discussions.
This work was supported by a FRIPRO grant from The Research Council of Norway (Project No. 345079).

 \section*{Author Declarations}
The authors have no conflicts to disclose.

\section*{Data availability}
All simulation scripts and the generated data will be made available.

\bibliographystyle{apsrev4-2}
\bibliography{references}

\clearpage
\renewcommand{\theequation}{{S}\arabic{equation}}\setcounter{equation}{0}
\renewcommand{\thefigure}{{S}\arabic{figure}}\setcounter{figure}{0} 
\renewcommand{\thesection}{{S}\arabic{section}}\setcounter{section}{0} 
\begin{widetext}

\section*{Supplemental Information for: ``Modeling electrolytes in nanopores by Monte Carlo simulations and the Bazant--Storey--Kornyshev model''}
\begin{center}
    Nader Nekoubin, David Fertig, Dezs\H{o} Boda, and Mathijs Janssen
\end{center}

\renewcommand{\thefigure}{S\arabic{figure}}

\section{Analytical solution to the linearized BSK equation}
\label{Ana_solution}

For small surface charges, the Boltzmann weights in \cref{eq:BSK} can be linearized, and the resulting equation has a general solution reading
\begin{align}
    \tilde{\psi}=A_1I_0(k_+\tilde{r})+A_2K_0(k_+\tilde{r})+A_3I_0(k_-\tilde{r})+A_4K_0(k_-\tilde{r}),
\end{align}
where $k_{\pm}^2=(1\pm\sqrt{1-4\delta^2})/(2\delta^2)$.

The boundary conditions \eqref{eq:BC_BSK3} and \eqref{eq:BC_BSK4} fix two integrations constant,  $A_2=A_4=0$, as the Bessel function of the second kind and its derivatives diverge at $\tilde{r}=0$.
Using \cref{eq:BC_BSK2}, we find
\begin{align}
    \delta A_1k_+^3I_1(k_+\zeta)+\delta A_3k_-^3I_1(k_-\zeta)+ A_1k_+^2I_0(k_+\zeta)+A_3k_-^2I_0(k_-\zeta)=0
\end{align}
which gives
\begin{align}\label{eq:A3}
    A_3=-A_1\dfrac{\delta k_+^3I_1(k_+\zeta)+k_+^2I_0(k_+\zeta)}{\delta k_-^3I_1(k_-\zeta)+k_-^2I_0(k_-\zeta)}=-A_1\Lambda,
\end{align}
where $\Lambda=\dfrac{\delta k_+^3I_1(k_+\zeta)+k_+^2I_0(k_+\zeta)}{\delta k_-^3I_1(k_-\zeta)+k_-^2I_0(k_-\zeta)}$.

Using \cref{eq:BC_BSK1_2,eq:A3}, and $1-\delta^2k_{\pm}^2=k_{\pm}^{-2}$, we find
\begin{align}
    A_1=\dfrac{\tilde{\sigma}\tilde{R}}{\zeta\left(\dfrac{I_1(k_+\zeta)}{k_+}-\Lambda \dfrac{I_1(k_-\zeta)}{k_-}\right)},
\end{align}
giving the potential in the diffuse layer,
\begin{align}\label{eq:analytical_solution}
    \tilde{\psi}=\dfrac{\tilde{\sigma}\tilde{R}k_+k_-[I_0(k_+\tilde{r})-\Lambda I_0(k_-\tilde{r})]}{\zeta\left[k_-I_1(k_+\zeta)-k_+\Lambda I_1(k_-\zeta)\right]}.
\end{align}
With this diffuse layer potential at hand, we find the Stern layer potential using \cref{eq:analytical solution}.

\section{Numerical implementation and validation of BSKB--Stern model}\label{numerical solution}

For the numerical implementation of the BSKB--Stern model, 
 we use $\tilde{w}=\partial^2_{\tilde{r}} \tilde{\psi}$ to split \cref{eq:BSK} into two parts,
\begin{subequations}\label{eq:BSK_numerical}
\begin{align}
    \label{eq:splitted_eq1}
    &\tilde{w}=\partial^2_{\tilde{r}} \tilde{\psi},\\
    \label{eq:splitted_eq2}
    &\delta^2\left(\partial^2_{\tilde{r}}\tilde{w}+\dfrac{2}{\tilde{r}}\partial_{\tilde{r}} \tilde{w}-\dfrac{1}{\tilde{r}^2} \tilde{w}+\dfrac{1}{\tilde{r}^3}\partial_{\tilde{r}} \tilde{\psi}\right) - \tilde{w}-\dfrac{1}{\tilde{r}}\partial_{\tilde{r}} \tilde{\psi} = \tilde{z}_+\exp(-z_+\tilde{\psi})+\tilde{z}_-\exp(-z_-\tilde{\psi}),
\end{align}
\end{subequations}
while the boundary conditions \eqref{eq:BC_BSK3}, and \eqref{eq:BC_BSK4}, \eqref{eq:BC_BSK1_2}, and \eqref{eq:BC_BSK2} turn into
\begin{subequations}\label{eq:BSK_BC_numerical}
\begin{align}  
    &\partial_{\tilde{r}}\tilde{\psi}\Big|_{\tilde{r}=0}=0\label{eq:BC_sp3} ,\\     
    &\partial_{\tilde{r}}w\Big|_{\tilde{r}=0}=0\,\label{eq:BC_sp4},\\
    &\partial_{\tilde{r}} \tilde{\psi}-\delta^2\left(\partial_{\tilde{r}}\tilde{w} + \dfrac{1}{\tilde{r}}\tilde{w} - \dfrac{1}{\tilde{r}^2}\partial_{\tilde{r}}\tilde{\psi}\right)\Big|_{\tilde{r}=\zeta^-}
    =\tilde{\sigma}\tilde{R}/\zeta\label{eq:BC_sp1},\\
    &\delta\left(\partial_{\tilde{r}}\tilde{w} + \dfrac{1}{\tilde{r}}\tilde{w} - \dfrac{1}{\tilde{r}^2}\partial_{\tilde{r}}\tilde{\psi}\right)\Big|_{\tilde{r}=\zeta^-}
    = -\tilde{w}-\dfrac{1}{\tilde{r}}\partial_{\tilde{r}}\tilde{\psi}\Big|_{\tilde{r}=\zeta^-}\label{eq:BC_sp2},
\end{align}
\end{subequations}
where \cref{eq:BC_sp3,eq:BC_sp1} are used to solve \cref{eq:splitted_eq1}, and \cref{eq:BC_sp2,eq:BC_sp4} are used to solve \cref{eq:splitted_eq2}.

We used a finite difference method based on the point-wise Gauss--Seidel method with central discretizations of both the equations and boundary conditions.
Two convergence criteria were considered: the maximum normalized differences between the calculated values ($\tilde{\psi}$ and $\tilde{w}$ at each node) of subsequent iterations, and the maximum normalized residuals of each equation at each node should be less than $10^{-8}$.

\Cref{fig:comparisonAnaNum}(a) shows numerical solutions to \cref{eq:BSK_numerical,eq:BSK_BC_numerical} for the potential $e\psi/kT$ along the radial coordinate, $r/\lambda_D$, for a small surface charge density $\tilde{\sigma}=-0.01$ (and other parameters listed in the figure caption).
The same panel also shows the analytical solution \cref{eq:analytical_solution} for the diffuse layer, extended in the Stern layer as explained below that equation.
We observe excellent agreement between the numerical and analytical results---their difference is shown in \cref{fig:comparisonAnaNum}(b)---confirming the accuracy of the developed numerical solver.

\Cref{fig:grid_independence} shows numerical results for the ionic charge density $q/(S e)$ profiles calculated for different grid sizes $(\Delta\tilde{r})$; all data correspond to a 3:1 electrolyte with a large non-dimensional correlation length ($\delta=5$) in a highly charged pore  ($\sigma=-3\,\mathrm{e/nm^2}$).
The maximum scaled difference between non-dimensional charge densities calculated by two last finest grids is less than $10^{-3}$, showing the grid independence of results.
The grid size of $\Delta\tilde{r}_0=0.01$ is employed in the preparation of all results presented in this study. 
\begin{figure}[h]
\includegraphics[width=0.5\linewidth]{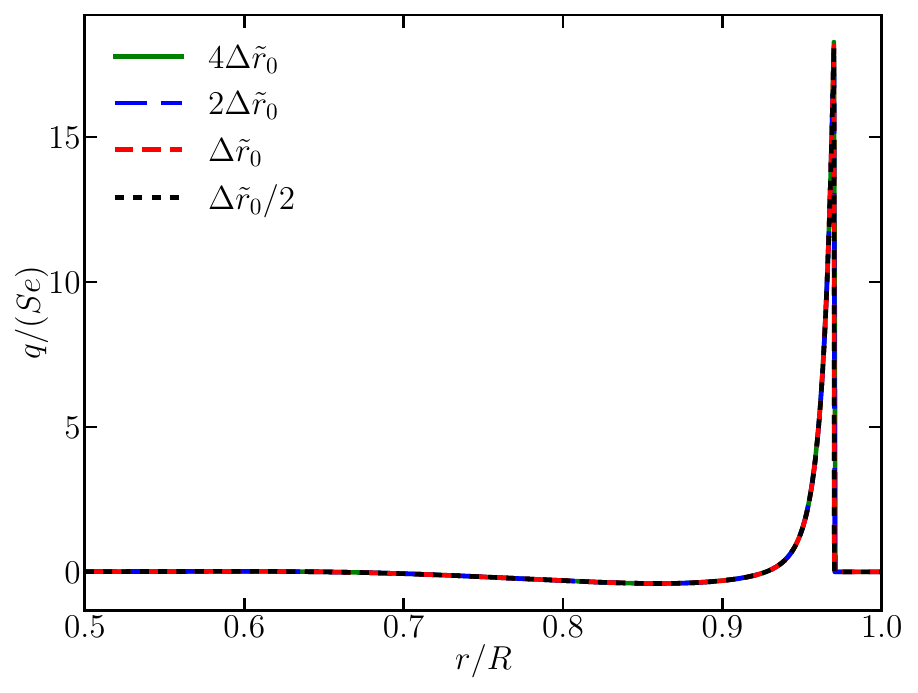}
\caption{\label{fig:grid_independence} Charge density profiles calculated by different grid sizes, for 3:1, $R=5 \,\mathrm{nm}$, $n_\infty=1\,\mathrm{M}, \sigma=-3\,\mathrm{e/nm^2}, \mathrm{and}\, \delta=5$, showing the grid independence of results.}
\end{figure}

\section{Charge density profiles for all considered cases}

We show MC charge density profiles for 1:1, 2:1, 3:1, and 2:2 electrolytes and surface charges densities $\sigma=0, -0.01, -0.03, -0.05, -0.1, -0.3, -0.5, -1, -3\,\mathrm{e/nm^2}$
for $n_\infty=0.1\,\mathrm{M}$ and $R=2\,\mathrm{nm}$ (\cref{fig:S_charge_density_1}); $n_\infty=1\,\mathrm{M}$ and $R=2\,\mathrm{nm}$ (\cref{fig:S_charge_density_2}); $n_\infty=0.1\,\mathrm{M}$ and $R=5\,\mathrm{nm}$ (\cref{fig:S_charge_density_3}); and $n_\infty=1\,\mathrm{M}$ and $R=5\,\mathrm{nm}$ (\cref{fig:S_charge_density_4}). 
For each parameter setting for which a BSKB--Stern fit converged, we show the model prediction, and report the corresponding $\delta$ fit parameter. 
Notably, the value  $\delta=3.66\times10^{-4}$ appears in many panels; this value follows from the parameters chosen for the Golden-Section search method, which is terminated when the width of search-interval becomes smaller than $0.001$, and the midpoint of final search-interval is taken as fitted $\delta$. So the repeated values of $3.66\times10^{-4}$ are basically equal to $0$. 

\begin{figure*}
    \centering
    \includegraphics[width=0.95\textwidth]{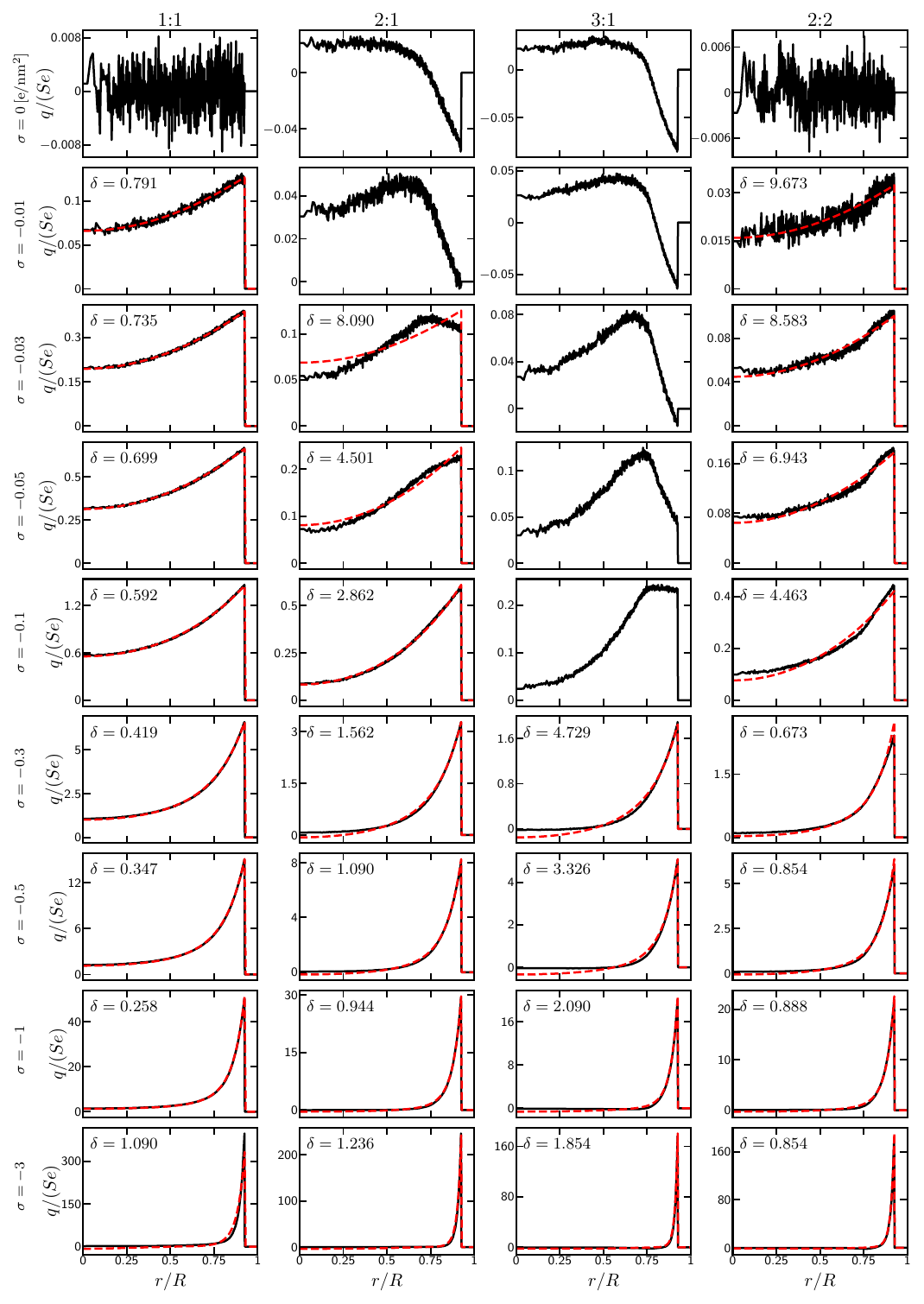}
    \caption{\label{fig:S_charge_density_1}
    Charge density profiles for 1:1, 2:1, 3:1, and 2:2 electrolytes at
    $n_\infty=0.1\,\mathrm{M}$ and $R=2\,\mathrm{nm}$ and various surface charges ($\sigma=0,-0.01, -0.03, -0.05, -0.1, -0.3, -0.5, -1, -3\,\mathrm{e/nm^2}$) as indicated, obtained by numerical (red dashed lines) and MC simulations (solid black lines).
    }
\end{figure*}

\begin{figure*}
    \centering
    \includegraphics[width=0.95\textwidth]{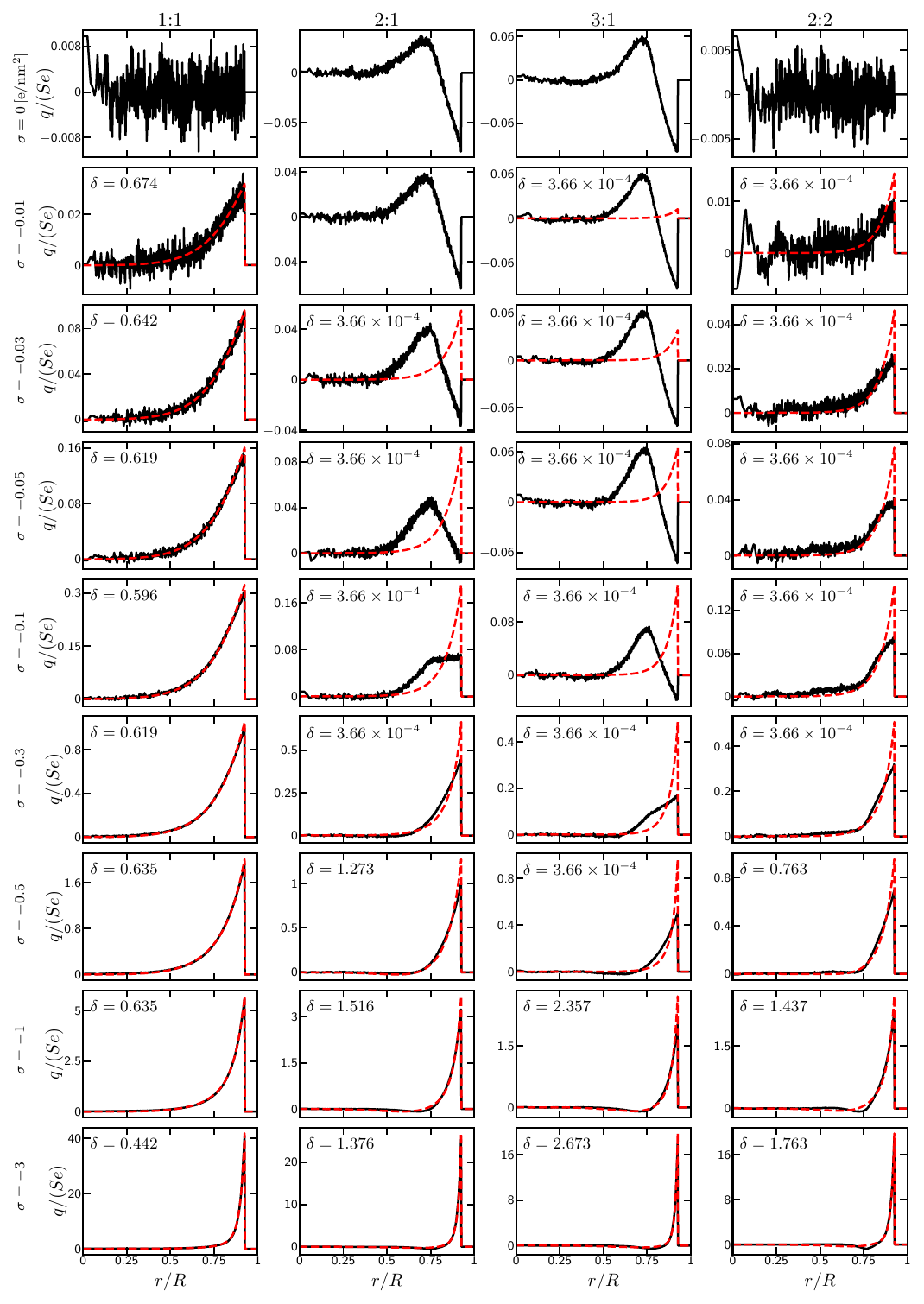}
    \caption{\label{fig:S_charge_density_2}
    Charge density profiles for 1:1, 2:1, 3:1, and 2:2 electrolytes at
    $n_\infty=1\,\mathrm{M}$ and $R=2\,\mathrm{nm}$ and various surface charges ($\sigma=0, -0.01, -0.03, -0.05, -0.1, -0.3, -0.5, -1, -3\,\mathrm{e/nm^2}$) as indicated, obtained by numerical (red dashed lines) and MC simulations (solid black lines).
    }
\end{figure*}

\begin{figure*}
    \centering
    \includegraphics[width=0.95\textwidth]{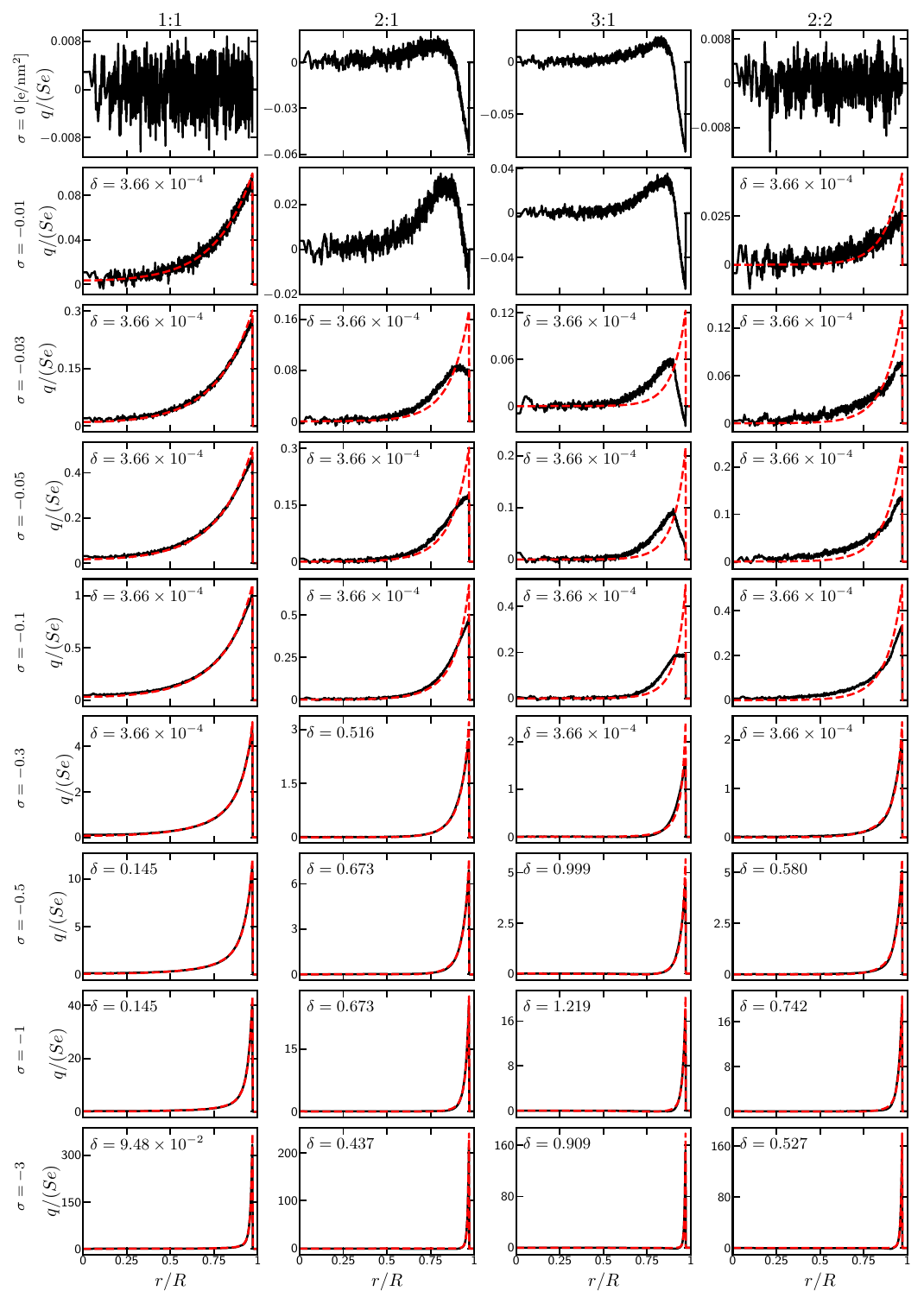}
    \caption{\label{fig:S_charge_density_3}
    Charge density profiles for 1:1, 2:1, 3:1, and 2:2 electrolytes at
    $n_\infty=0.1\,\mathrm{M}$ and $R=5\,\mathrm{nm}$ and various surface charges ($\sigma=0, -0.01, -0.03, -0.05, -0.1, -0.3, -0.5, -1, -3\,\mathrm{e/nm^2}$) as indicated, obtained by numerical (red dashed lines) and MC simulations (solid black lines).
    }
\end{figure*}

\begin{figure*}
    \centering
    \includegraphics[width=0.95\textwidth]{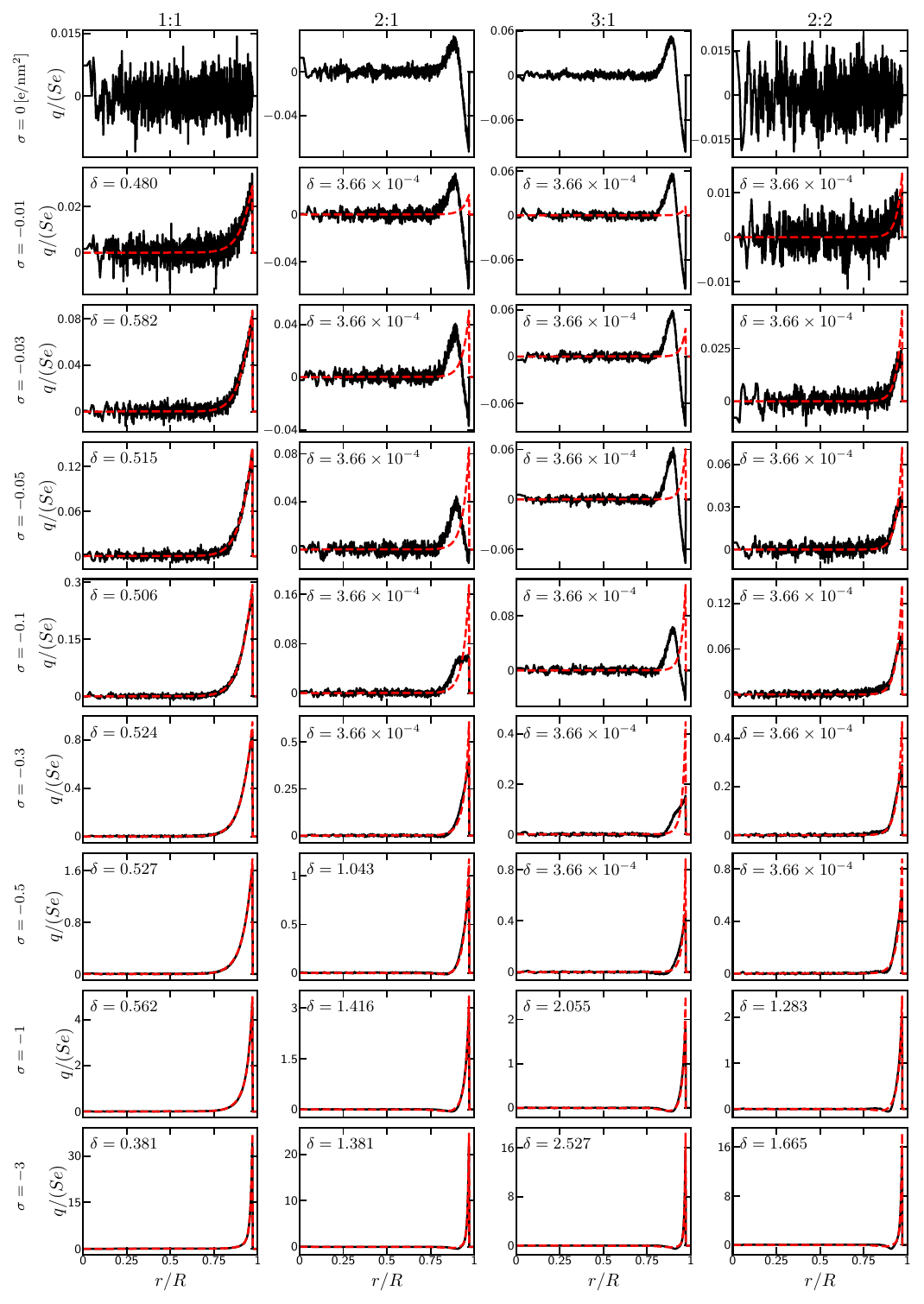}
    \caption{\label{fig:S_charge_density_4}
    Charge density profiles for 1:1, 2:1, 3:1, and 2:2 electrolytes at
    $n_\infty=1\,\mathrm{M}$ and $R=5\,\mathrm{nm}$ and various surface charges ($\sigma=0, -0.01, -0.03, -0.05, -0.1, -0.3, -0.5, -1, -3\,\mathrm{e/nm^2}$) as indicated, obtained by numerical (red dashed lines) and MC simulations (solid black lines).
    }
\end{figure*}

\section{Fit parameter $\delta$ and quality of fitting}
\label{fitting}
In \cref{fig:delta_charge}, the values of $\delta$ fitted by the Golden-Section search method are presented for all cases. The magnitudes of normalized root mean square (NRMS), normalized with the maximum absolute value of $q/(Se)$ for each case, are also shown with vertical bars. Vertical bars extend with the magnitude of NRMS in each direction at each point ($\pm \mathrm{NRMS}$). So, the heights of them are two times of NRMS for the sake of visualization, showing the fitting quality in each point.

\begin{figure}[h]
\includegraphics[width=0.95\linewidth]{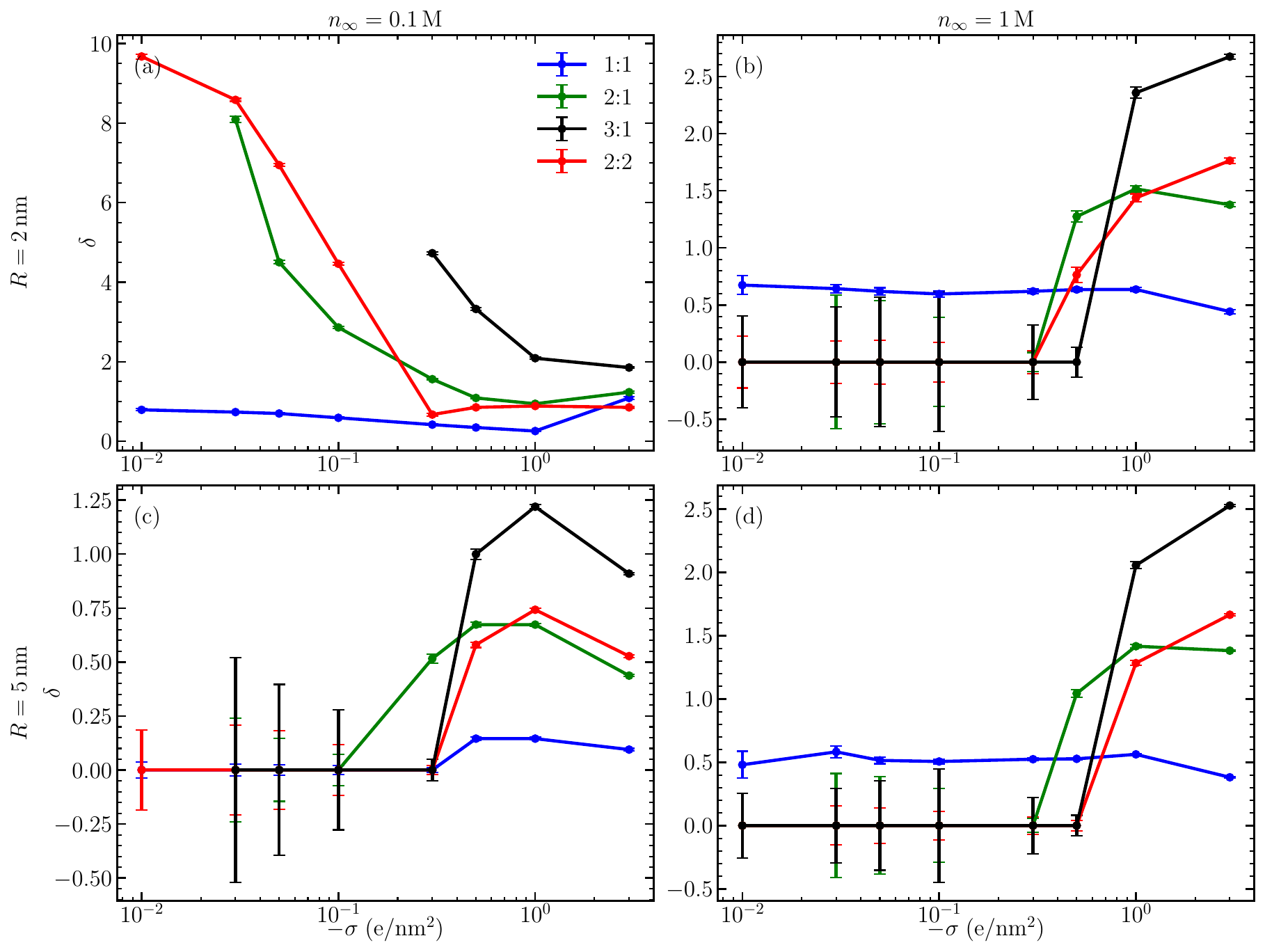}
\caption{\label{fig:delta_charge} Calculated values of $\delta$ for all cases, where the vertical bars at each point indicate two times of NRMS.}
\end{figure}

\end{widetext}
\end{document}